\documentclass[AMA,STIX1COL]{WileyNJD-v2}
\usepackage{moreverb,pgfplots,pgfplotstable}
  \usetikzlibrary{trees,arrows,decorations}
  \tikzset{
    declare function={
      normcdf(\x)=1/(1 + exp(-0.07056*((\x)/1)^3 - 1.5976*(\x)/1));
    }
  }
  \definecolor{v6}{RGB}{0,122,150}
  \definecolor{v3}{RGB}{63,58,114}
  \definecolor{v10}{RGB}{0,190,139}

\newcommand{\no}{n_{\text{obs}}}
\newcommand{\ns}{n_{\text{sim}}}

\articletype{Tutorial in Biostatistics}

\received{29 November 2017}
\revised{23 August 2018}
\accepted{02 October 2018}

\begin{document} \frenchspacing

\title{Using simulation studies to evaluate statistical methods}

\author[1]{Tim P Morris*}
\author[1]{Ian R White}
\author[2]{Michael J Crowther}
\authormark{TP Morris, IR White, MJ Crowther}

\address[1]{\orgdiv{London Hub for Trials Methodology Research}, \orgname{MRC Clinical Trials Unit at UCL}, \orgaddress{\state{London}, \country{UK}}}
\address[2]{\orgdiv{Biostatistics Research Group, Department of Health Sciences}, \orgname{University of Leicester}, \orgaddress{\state{Leicester}, \country{UK}}}
\corres{*Tim P Morris, MRC Clinical Trials Unit at UCL. \email{tim.morris@ucl.ac.uk}}
\presentaddress{90 High Holborn, London WC1V 6LJ, UK}

\abstract[Abstract]{
  \noindent Simulation studies are computer experiments that involve creating data by pseudo-random sampling. A key strength of simulation studies is the ability to understand the behaviour of statistical methods because some `truth' (usually some parameter/s of interest) is known from the process of generating the data. This allows us to consider properties of methods, such as bias. While widely used, simulation studies are often poorly designed, analysed and reported. This tutorial outlines the rationale for using simulation studies and offers guidance for design, execution, analysis, reporting and presentation. In particular, this tutorial provides: a structured approach for planning and reporting simulation studies, which involves defining aims, data-generating mechanisms, estimands, methods and performance measures (`ADEMP'); coherent terminology for simulation studies; guidance on coding simulation studies; a critical discussion of key performance measures and their estimation; guidance on structuring tabular and graphical presentation of results; and new graphical presentations. With a view to describing recent practice, we review 100 articles taken from Volume~34 of \textit{Statistics in Medicine} that included at least one simulation study and identify areas for improvement.
}
\keywords{Simulation studies; Monte Carlo; Simulation design; Simulation reporting; Graphics for simulation}

\maketitle

\section{Introduction} \label{s:intro}
Simulation studies are computer experiments that involve creating data by pseudo-random sampling from known probability distributions. They are an invaluable tool for statistical research, particularly for the evaluation of new methods and for the comparison of alternative methods. Simulation studies are much used in the pages of \textit{Statistics in Medicine}, but our experience is that some statisticians lack the necessary understanding to execute a simulation study with confidence, while others are over-confident and so fail to think carefully about design and report results poorly. Proper understanding of simulation studies would enable the former to both run and critically appraise published simulation studies themselves and the latter to conduct simulation studies with greater care and report with transparency. Simulation studies are empirical experiments and statisticians should therefore use knowledge of experimental design and analysis in running them. As we shall see, inadequacies with design, analysis and reporting lead to uncritical use and interpretation of simulation studies. In this context, better understanding of the rationale, design, execution, analysis and reporting of simulation studies is necessary to improve understanding and interpretation of the findings.

Simulation studies are used to obtain empirical results about the performance of statistical methods in certain scenarios, as opposed to more general analytic (algebraic) results, which may cover many scenarios. It is not always possible, or may be difficult, to obtain analytic results. Simulation studies come into their own when methods make wrong assumptions or data are messy because they can assess the resilience of methods in such situations. This is not always possible with analytic results, where results may apply only when data arise from a specific model.

`Monte Carlo simulation' means statistical techniques that use pseudo-random sampling, and has many uses that are not simulation \textit{studies}. For example, it is required to implement multiple imputation and Markov Chain Monte Carlo methods. The remainder of this paper does not consider these uses, unless the properties of some such method are being evaluated by a simulation study.

There are many ways to use simulation studies in medical statistics. Some examples are:
\begin{list}{$\circ$}{\leftmargin=1em\labelwidth=1em\topsep=0em\parsep=0em\itemsep=0em}
  \item To check algebra (and code), or to provide reassurance that no large error has been made, where a new statistical method has been derived mathematically.
  \item To assess the relevance of large-sample theory approximations (e.g. considering the sampling distribution of an estimator) in finite samples.
  \item For the absolute evaluation of a new or existing statistical method. Often a new method is checked using simulation to ensure it works in the scenarios for which it was designed.
  \item For comparative evaluation of two or more statistical methods.
  \item For calculation of sample size or power when designing a study under certain assumptions\cite{feiveson02}.
\end{list}
This article is focused primarily on using simulation studies for the evaluation of methods. Simulation studies for this purpose are typically motivated by frequentist theory and used to evaluate the frequentist properties of methods, even if the methods are Bayesian\cite{rubin84,grieve16}.

It seems that as a profession we fail to follow good practice regarding design, analysis, presentation and reporting in our simulation studies, as lamented previously by Hoaglin \& Andrews\cite{hoaglin75}, Hauck \& Anderson\cite{hauck84}, Ripley\cite{Ripley87}, Burton \textit{et al.}\cite{Burton06}, and Koehler, Brown \& Haneuse\cite{Koehler++09}. For example, few reports of simulation studies acknowledge that Monte Carlo procedures will give different results when based on a different set of random numbers and hence are subject to uncertainty, yet failing to report measures of uncertainty would be unacceptable in medical research.

There exist some wonderful books on simulation methods in general \cite{Ripley87,morgan84,chang11} and several excellent articles encouraging rigour in specific aspects of simulation studies (for example \cite{feiveson02, hoaglin75, hauck84, Koehler++09, diazemp02, White10simsum, smith11, crowther13, holford99, okelly17}) but, until now, no unified practical guidance on simulation studies. This tutorial provides such guidance. More specifically we: introduce a structured approach for planning and reporting simulation studies; provide coherent terminology for simulation studies; offer guidance on coding simulation studies; critically discuss key performance measures and their estimation; make suggestions for structuring tabular and graphical presentation of results; and introduce several new graphical presentations. This guidance should enable practitioners to execute a simulation study for the first time and contains much for more experience practitioners. For reference, the main steps involved, key decisions and recommendations are summarised in table \ref{t:recom}.

The structure of this tutorial is as follows. We describe a review of a sample of the simulation studies reported in \textit{Statistics in Medicine} Volume~34 (section \ref{s:review}). In section \ref{s:planning} we outline a systematic approach to planning simulation studies, using the new `\textsc{ademp}' structure (which we define there). Section \ref{s:coding} gives guidance on computational considerations for coding simulation studies. In section \ref{s:analysis}, we discuss the purposes of various performance measures and their estimation, stressing the importance of estimating and reporting the Monte Carlo standard error (SE) as a measure of uncertainty due to using a finite number of simulation repetitions. Section \ref{s:report} outlines how to report simulation studies, again using the \textsc{ademp} structure, and offers guidance on tabular and graphical presentation of results. Section \ref{s:egsims} works through a simple simulation to illustrate in practice the approaches that we are advocating. Section \ref{s:conclusions} offers some concluding remarks, with a short subsection (\ref{s:future}) that considers some future directions. Examples are drawn from the review and from the authors' areas of interest (which relate mainly to modeling survival data, missing data, meta-analysis and randomised trial design).

\begin{table} \caption{Key steps and decisions in the planning, coding, analysis and reporting of simulation studies\label{t:recom}} \centering
  \begin{tabular} {c p{14cm} r}
  \hline
   & & Section \\
  \multicolumn{2}{l}{\textsc{Planning}} & \ref{s:planning} \\
  \multicolumn{2}{l}{Aims} & \ref{ss:aims} \\
  $\cdot$ & Identify \textit{specific} aims of simulation study. & \\
  \multicolumn{2}{l}{Data-generating mechanisms} & \ref{ss:dgm} \\
  $\cdot$ & In relation to the aims, decide whether to use resampling or simulation from some parametric model. & \\
  $\cdot$ & For simulation from a parametric model, decide how simple or complex the model should be and whether it should be based on real data. & \\
  $\cdot$ & Determine what factors to vary and the levels of factors to use. & \\
  $\cdot$ & Decide whether factors should be varied fully factorially, partly factorially or one-at-a-time. & \\
  \multicolumn{2}{l}{Estimand / target of analysis} & \ref{ss:est} \\
  $\cdot$ & Define estimands and/or other targets of the simulation study. &  \\
  \multicolumn{2}{l}{Methods} & \ref{ss:meth} \\
  $\cdot$ & Identify methods to be evaluated and consider whether they are appropriate for estimand/target identified. For method comparison studies, make a careful review of the literature to ensure inclusion of relevant methods. & \\
  \multicolumn{2}{l}{Performance measures} & \ref{ss:pm},~\ref{ss:pms} \\
  $\cdot$ & List all performance measures to be estimated, justifying their relevance to estimands or other targets. & \\
  $\cdot$ & For less-used performance measures, give explicit formulae for the avoidance of ambiguity. & \ref{ss:pms} \\
  $\cdot$ & Choose a value of $\ns$ that achieves acceptable Monte Carlo SE for key performance measures. & \ref{ss:pms}, \ref{ss:sampsi} \\
  \hline
  \multicolumn{2}{l}{\textsc{Coding and execution}} & \ref{s:coding} \\
  $\cdot$ & Separate scripts used to analyse simulated datasets from scripts to analyse estimates datasets. &  \\
  $\cdot$ & Start small and build up code, including plenty of checks. &  \\
  $\cdot$ & Set the random number seed once per simulation repetition. &  \\
  $\cdot$ & Store the random number states at the start of each repetition. &  \\
  $\cdot$ & If running chunks of the simulation in parallel, use separate streams of random numbers \cite{haramoto08}. &  \\
  \hline
  \multicolumn{2}{l}{\textsc{Analysis}} & \ref{s:analysis} \\
  $\cdot$ & Conduct exploratory analysis of results, particularly graphical exploration. &  \\
  $\cdot$ & Compute estimates of performance and Monte Carlo SEs for these estimates.  & \ref{ss:pms} \\
  \hline
  \multicolumn{2}{l}{\textsc{Reporting}} & \ref{s:report} \\
  $\cdot$ & Describe simulation study using \textsc{ademp} structure with sufficient rationale for choices. &  \\
  $\cdot$ & Structure graphical and tabular presentations to place performance of competing methods side-by-side. &  \\
  $\cdot$ & Include Monte Carlo SE as an estimate of simulation uncertainty. & \ref{ss:pms} \\
  $\cdot$ & Publish code to execute the simulation study including user-written routines. & \ref{s:conclusions} \\
  \hline
  \end{tabular}
\end{table}

\section{Simulation in practice: a review of \textit{Statistics in Medicine}, Volume~34} \label{s:review}

We undertook a review of practice based on articles published in Volume~34 of \textit{Statistics in Medicine}~(2015). This review recorded information relevant to the ideas in this article. In this section we briefly outline the review but do not give results, which instead are provided at relevant points. The raw data on which results are based are provided as a Stata file in the supplementary materials (pared down, without comments). 

We restricted attention to research articles, excluding tutorials in biostatistics, commentaries, book reviews, corrections, letters to the editor and authors' responses. In the volume, there were a total of 264 research articles of which 199~(75\%) included at least one simulation study.

In planning the review, we needed to select a sample size. Most of the questions of interest involved binary answers. For such questions, to estimate proportions with maximum standard error of 0.05 (occurring when the proportion is 0.5), we randomly selected 100 articles that involved a simulation study, before randomly assigning articles to a reviewer. TPM reviewed 35 simulation studies, IRW reviewed 34 and MJC reviewed 31. In case the reviewer was an author or co-author of the article, the simulation study was swapped with another reviewer. TPM also reviewed five of the simulation studies allocated to each of the other reviewers to check agreement on key information (results on agreement are given in appendix \ref{app:dgms} and figure \ref{f:agree}).

\section{Planning simulation studies using \textsc{ademp}} \label{s:planning}

For clarity about the concepts that will follow, we introduce some notation in table \ref{t:notation}. Note that $\theta$ is used to represent a connceptual estimand and its true value.

\begin{table}
  \caption{Description of notation\label{t:notation}}
  \centering
  \begin{tabular}{rl}
     \hline
     $\theta$ & An estimand (conceptually); also true value of the estimand \\
     $\no$ & Sample size of a simulated dataset \\
     $\ns$ & Number of repetitions used; the simulation sample size \\
     $i=1,\ldots ,\ns$ & Indexes the repetitions of the simulation \\
     $\hat{\theta}$ & the estimator of $\theta$ \\
     $\hat{\theta}_i$ & the estimate of $\theta$ from the $i$th repetition \\
     $\bar{\theta}$ & the mean of $\hat{\theta}_i$ across repetitions \\
     $\text{Var}(\hat{\theta})$ & the true variance of $\hat{\theta}$, which can be estimated with large $\ns$ \\
     $\widehat{\text{Var}}(\hat{\theta}_i)$ & an estimate of $\text{Var}(\hat{\theta})$ from the $i$th repetition \\
     $\alpha$ & the nominal significance level \\
     $p_i$ & the p-value returned by the $i$th repetition \\
     \hline
   \end{tabular}
\end{table}

In the following sections, we outline the \textsc{ademp} structured approach to planning simulation studies. This acronym comes from: \textit{Aims}, \textit{Data-generating mechanisms}, \textit{Methods}, \textit{Estimands}, \textit{Performance measures}.

\subsection{Aims} \label{ss:aims}
In considering the aims of a simulation study it is instructive to first consider desirable properties of an estimator $\hat{\theta}$ from a frequentist perspective.

\begin{list}{$\circ$}{\leftmargin=2em\labelwidth=2em\topsep=0em\parsep=0em\itemsep=0em}
  \item[1.] $\hat{\theta}$ should be consistent: as $n \to \infty$, $\hat{\theta} \to \theta$.
  It is also desirable that $\hat{\theta}$ be unbiased for $\theta$ in finite samples: $\text{E}(\hat{\theta})=\theta$ (though arguably less important since unbiasedness is not an invariant property). Some estimators may be consistent but exhibit small-sample bias (logistic regression for example). 
  \item[2.] The sample estimate $\widehat{\text{Var}}(\hat{\theta})$ should be a consistent estimate of the sampling variance of $\hat{\theta}$ (see for example \cite{KenwardRoger97}).
  \item[3.] Confidence intervals should have the property that at least $100(1-\alpha)\%$ of intervals contain $\theta$ (see section \ref{ss:pms}).
  \item[4.] It is desirable that $\text{Var}\hat{(\theta)}$ be as small as possible: that $\hat{\theta}$ be an efficient estimator of $\theta$.
\end{list}
There are other properties we might desire, but these tend to involve combinations of the above. For example, short average confidence interval length may be desirable; this relates to (4) and its validity depends on (1), (2) and (3). Mean squared error is a combination of (1) and (4). Further, properties may be traded off; small bias may be accepted if there is a substantial reduction in $\text{Var}\hat{(\theta)}$.

The aims of a simulation study will typically be set out in relation to the above properties, depending on what specifically we wish to learn. A simulation study might primarily investigate: large- or small-sample bias (e.g.\ \cite{white97}; precision, particularly relative to other available methods (e.g.\ \cite{WhiteThompson05}); Variance estimation (e.g.\ \cite{hughes14b}); or robustness to misspecification (e.g.\ \cite{morris14pmm}).

There is a distinction between simulation studies that offer a proof-of-concept, \textit{i.e.}\ showing that a method is viable (or fallible) in some settings, and those that aim to stretch or break methods, \textit{i.e.}\ identifying settings where the method may fail. Both are useful and important in statistical research. For example, one may be faced with two competing methods of analysis, both of which are equally easy to implement. Even if the choice is unlikely to materially affect the results, it may be useful to have unrealistically extreme data-generating mechanisms to understand when and how each method fails; e.g.\ \cite{morris14pmm}.

Alternatively, it may be of interest to compare methods where some or all methods have been shown to work in principle but the methods under scrutiny were designed to address slightly different problems. They may be put head-to-head in realistic scenarios. This could be to investigate properties when one method is correct -- \textit{How badly do others fail?} -- or when all are incorrect in some way -- \textit{Which is most robust?} No method will be perfect and it is useful to understand how methods are likely to perform in the sort of scenarios that might be expected in practice. However, such an approach poses tough questions in terms of generating data: \textit{Does the data-generating mechanism favour certain methods over others?} \textit{How can this be checked and justified?} One common justification is by reference to motivating data. However, in the absence of a broad spectrum of such motivating data, there is a risk of failing to convince readers that a method is fit for general use.

\subsection{Data-generating mechanisms} \label{ss:dgm}

We use the term `data-generating mechanism' to denote how random numbers are used to generate a dataset. This is in preference to `data-generating \textit{model}', which implies parametric models and so is a specific class of data-generating mechanism. It is not the purpose of this article to explain how specific types of data should be generated. See Ripley~\cite{Ripley87} or Morgan~\cite{morgan84} for methods to simulate data from specific distributions. In planning a simulation study, it is usual to spend more time deciding on data-generating mechanisms than any other element of \textsc{ademp}. There are many subtleties and potential pitfalls, some of which we will mention below.

Data may be generated by producing parametric draws from a known model (once or many times), or by repeated resampling with replacement from a specific dataset (where the true data-generating model is unknown). For resampling studies, the true data-generating mechanism is unknown and resamples are used to study the sampling distribution. While parametric simulation can explore many different data-generating mechanisms (which may be completely unrealistic), resampling typically explores only one mechanism (which will be relevant for at least the study at hand). 

The choice of data-generating mechanism(s) will depend on the aims. As noted above, we might investigate a method under a simple data-generating mechanism, a realistic mechanism, or a completely unrealistic mechanism designed to stretch a method to breaking point.

Simulation studies provide us with empirical results for specific scenarios. For this reason, simulation studies will often involve more than one data-generating mechanism to ensure coverage of different scenarios. For example, it is very common to vary the sample size of simulated datasets because performance often varies over $\no$ (see section \ref{ss:pms} and figure~\ref{f:mse}).

Much can be controlled in a simulation study and statistical principles for designing experiments therefore can and should be called on. In particular, there is often more than one factor that will vary across specific data-generating mechanisms. Factors that are frequently varied are sample size (several values) and true parameter values (for example, setting one or more parameters to be zero or non-zero). Varying these factorially is likely to be more informative than one-by-one away from a `base-case' data-generating mechanism, as doing so permits the exploration of interactions between factors. There are however practical implications that might make this infeasible. The first regards presentation of results (covered in section \ref{s:report}) and the second computational time. If the issue is simply around presentation, it may be preferable to define a `base case' but perform a factorial simulation study anyway and, if results are consistent with no interaction, presentation can vary factors away from the base case one-by-one.

If the main issue with executing a fully factorial design is computational time, it may be necessary for the simulation study to follow a non-factorial structure. Three approaches are noted below.

A first pragmatic check may be to consider interactions only where main effects exist. If performance seems acceptable and does not vary according to factor A, it would seem unlikely to have chosen a data-generating mechanism that happened to exhibit this property when performance would have been poor for other choices of data-generating mechanism.

A more careful approach could be taken based on making and checking predictions beyond the data-generating mechanisms initially used; an idea similar to external validation. Suppose we have two factors, A and B, where $\text{A} \in \{1, \ldots , 8\}$ and $\text{B} \in \{1, \ldots , 5\}$ in the data-generating mechanism. The base-case is $\text{A}=1,\text{B}=1$. If the non-factorial portion of the design varies A from 1~to~8 holding $\text{B}=1$, and varies B from 1~to~5 holding $\text{A}=1$, this portion of the simulation study could be used to predict performance when $\text{A}=8,\text{B}=5$. Predictions may be purely qualitative (`bias increases as A increases and as B increases so when we increase both together we would expect even larger bias'), or quantitative (based on the marginal effects after fitting a model to existing results, thereby producing explicit predictions at unexplored values of A and B). The simulation study can then be re-run for that single data-generating mechanism, say $\text{A}=8, \text{B}=5$ and predictions compared with the empirical results (with a responsibility to expore further when predictions are poor or incorrect).

Finally, a more satisfactory solution is of course to use a fractional factorial design for the data-generating mechanisms\cite{skrondal00,grieve16}.

We now issue some specific pitfalls to help readers in choosing data-generating mechanisms (specifically acknowledging Stephen Senn's input).
\begin{enumerate}
  \item Resampling with replacement from a dataset but failing to appreciate that results are relevant to an infinite population with the exact characteristics of that dataset. For example, if a trial had a non-significant result, the treatment effect is non-zero in the implicit population\cite{graves18}.
  \item Missing the distinction between the logical flow of Bayesian and frequentist simulation. Repeated simulation with a single parameter value is explicitly frequentist. The fact that $\hat{\theta}$ is on average equal to $\theta$ does not imply that $\theta$ is on average equal to $\hat{\theta}$.
  \item Failing to distinguish between what the simulator can know and what the estimator can know\cite{sennletter94}.
  \item Employing tricks in data-generation without appreciating that the resulting data are not what was desired. As an example, suppose one wishes to simulate bivariate data with a desired $R^2$, say 0.3. For any given repetition, the observed $R^2$ will not equal $0.3$, but this could be fixed by scaling the residuals. This would produce unintended side effects for other statistics.
\end{enumerate}

In our review, 97 simulation studies used some form of parametric model to generate data while three used resampling methods. Of the 97 that simulated from a parametric model, 27 based parameter values on data, one based parameter values partly on data, and the remaining 69 on no data. Of these 97, 91~(94\%) provided the parameters used. The most careful example \cite{kuss15} explored analysis of meta-analysis data and drew the design factors from empirical data on 14,886 performed meta-analyses from 1,991 Cochrane Reviews. The total number of data-generating mechanisms per simulation study ranged from 1 to $4.2 \times 10^{10}$; figure \ref{f:dgms} (in the appendix) summarises aspects of the data-generating mechanisms. Where more than one factor was varied, fully factorial designs were the most frequent, while some used partially factorial designs. None used any of the alternative approaches we have described.

\subsection{Estimands and other targets} \label{ss:est}

The majority of simulation studies evaluate or compare methods for estimating one or more population quantities, which we term \textit{estimands} and denote by $\theta$. An estimand is usually a parameter of the data generating model, but is occasionally some other quantity. For example, when fitting regression models with parameter $\mathbf{\beta} = (\beta_0 \ldots \beta_c)$, the estimand may be a specific $\beta$, a measure of prognostic ability, the fitted outcome mean, or something else. In order to choose a relevant estimand, it is important to understand the aims of analysis in practice.

The choice of estimand is sometimes a simple matter of stating a parameter of interest. At other times, it is more subtle. For example, a logistic regression model unadjusted for covariates implies a marginal estimand; a model adjusted for covariates implied a conditional estimand with a different true value (this example is expanded on in section \ref{ss:meth}). 

Not all simulation studies evaluate or compare methods that concern an estimand. Other simulation studies evaluate methods for testing a null hypothesis, for selecting a model, or for prediction. We refer to these as \textit{targets} of the simulation study. The same statistical method could be evaluated against multiple targets. For example, the best method to select a regression model to estimate the coefficient of an exposure (targeting an estimand) may differ from the best model for prediction of outcomes (targeting prediction). Where a simulation study evaluates methods for design, rather than analysis, of a biomedical study, the design is the target.

Table \ref{t:targets} summarises different possible targets of a simulation study and suggests some performance measures (described more fully in section \ref{ss:pm}) that may be relevant for each target, with examples taken from Volume~34.

\begin{table} \caption{Possible targets of a simulation study and relevant performance measures} \label{t:targets} \centering
\begin{tabular}{l l p{4cm} p{7cm}}
  \hline
  Statistical task & Target & Examples of performance & Example \\
  & & measures & \\
  \hline
  \textit{Analysis} & & & \\
  Estimation & Estimand & Bias, empirical SE, mean-squared error, coverage & Kuss compares a number of existing methods in terms of bias,  power and coverage \cite{kuss15} \\
  Testing & Null hypothesis & Type~I error rate, power & Chaurasia and Harel compare new methods inn terms of type~I and II error rates \cite{chaurasia15} \\
  Model selection & Model & Correct model rate, sensitivity or specificity for covariate selection & Wu \textit{et al.} compare four new methods in terms of `true positive' and `false positive' rates of covariate selection \cite{wu15} \\
  Prediction & Prediction/s & Measures of predictive accuracy, calibration, discrimination & Ferrante compares four methods in terms of mean absolute prediction error, etc. \cite{ferrante15} \\
  \hline
  \textit{Design} & & & \\
  Design a study & Selected design & Sample size, expected sample size, power / precision & Zhang compares designs across multiple data-generating mechanisms in terms of number of significant test results (described as `gain') and frequency of achieving the (near) optimal design \cite{zhang15}\\
  \hline
\end{tabular}
\end{table}

In our review, 64 simulation studies targeted an estimand, 21 targeted a null hypothesis, eight targeted a selected model, three targeted predictive performance, and four had some other target. Of the 64 targeting an estimand, 51 stated what the estimand was (either in the description of the simulation study or elsewhere in the article). A figure detailing the number of estimands in simulation studies that targeted an estimand is given in the appendix, figure~\ref{f:em}.

\subsection{Methods} \label{ss:meth}

The term `method' is generic. Most often it refers to a model for analysis, but might refer to a design or some procedure (such as a decision rule). For example, \cite{kahan13factor} and \cite{campbell12} evaluated procedures that involved choosing an analysis based on the result of a preliminary test in the same data.

In some simulation studies there will be only one method with no comparators. In this case, selecting the method to be evaluated is very simple. When we aim to compare several methods in order to identify the best, it is important to include serious contenders. There are two issues.

First, it is necessary to have knowledge of previous work in the area to understand which methods are and are not serious contenders. Some methods may be legitimately excluded if they have already been shown to be flawed, and it may be unnecessary to include such methods if the only consequences are repetition of previous research and bloating of results. An exception might be if a flawed method is used frequently in practice (for example \textit{last observation carried forward} with incomplete longitudinal data, or the `$3+3$'~design for dose-escalation).

Second, code. New methods are sometimes published but not implemented in any software (for example \cite{RobinsWang00} and \cite{reiter08}), implemented poorly, or implemented in unfamiliar software. For methods that have not been implemented, it is hard to argue that they should be included in simulation studies. Although R and Stata appear to dominate for user-written methods, it is not uncommon to see methods implemented in other packages. See section \ref{ss:software} for a discussion of the situation where the methods under consideration are not all implemented in one same package. Note that one important role of simulation is to verify that code is correct.

Methods may also be excluded if they do not target the specified estimand/s. Understanding whether methods target an estimand or not can be subtle. Returning to the example of randomised trial with a binary outcome, one might compare two logistic regression analyses, one unadjusted and one adjusted for a covariate, where the estimand is the log odds ratio for randomised group. In a simulation study, one would be likely to find that the two methods give different mean estimates, and it would be tempting to conclude that at least one of the methods is biased. However, the two methods target different estimands -- that is, the true unadjusted and adjusted log odds ratios differ\cite{Hauck98}.

One way to tackle the problem of different estimands is to ensure that both methods estimate the same estimand: in the example of the randomised trial using logistic regression, this would involve post-processing the adjusted regression results to estimate the adjusted marginal odds ratio, which is the same estimand as the unadjusted analysis~\cite{zhang08}. This of course implies that the adjustment \textit{vs.}\ non adjustment is the method comparison we are interested in (it may not be), and that the conditional estimand is simply a nuisance part of standard adjustment. An alternative (but coarser) way to tackle the problem is to target the null hypothesis, if the two methods test the same null. In the logistic regression example described above, because the setting is a randomised trial, the null hypothesis  that the odds ratio equals 1 is the same whether the odds ratio is conditional or marginal.

The number of methods evaluated in our review of Volume~34 ranged from 1 to 33 (see figure \ref{f:em}). 

Non-convergence and other related issues such as perfect prediction (`separation'\cite{smeden16}) can blight some simulation studies. In such situations, there is a conceptual issue with defining a method. A `pure' method evaluation would simply assess performance of a model when it converges. However, in practice, an analyst whose model fails to converge would not give up but explore other models until one converges. Thus, evaluation of such a \textit{procedure} may be of interest in simulation studies. Crowther, Look and Riley evaluated such a procedure\cite{crowther14}: if a model failed to converge, they applied a model with more quadrature points. We will commment further on this issue in section \ref{ss:pms}.

\subsection{Performance measures} \label{ss:pm}

The term `performance measure' describes a numerical quantity used to assess the performance of a method. The equivalent term `operating characteristic' is sometimes used, particularly in the context of study designs (see for example \cite{royston11}). Statistical methods for estimation may output for example an estimate $\hat{\theta}_i$, an estimate of variance $\widehat{\text{Var}}(\hat{\theta})_i$ (or standard error $\widehat{\text{SE}}(\hat{\theta})_i$), degrees of freedom, confidence intervals, test statistics and more (such as an estimate of prognostic performance).

The performance measures required in a simulation study depend on the aims and what the study targets (see section \ref{ss:est}). When the target is an estimand, the most obvious performance measure to consider is bias: the amount by which $\hat{\theta}$ exceeds $\theta$ on average (this can be positive or negative). Precision and coverage of $(1-\alpha)$ confidence intervals will also be of interest. Meanwhile, if the target is a null hypothesis, power and type I error rates will be of primary interest. A simulation study targeting an estimand may of course also assess power and type~I error.

The performance measures seen in our review are summarised in table \ref{t:rpms}. The denominator changes according across performance measures because some are not applicable for some simulation studies. Further, sometimes simulation studies had secondary targets. For example, nine simulation studies primarily targeted a null hypothesis but secondarily targeted an estimand and could have assessed bias, and one of these did so. For eight articles, some performance measures were unclear. In some, a performance measure was given a name that its formula demonstrated to be misleading (an example is the term `mean error', which is bias, when the formula is for mean \textit{absolute} error), emphasising the importance of clear terminology in simulation studies.

\begin{table} \centering
\caption{Performance measures evaluated in review of Volume~34 (Frequency (and \%)) \label{t:rpms}}
\begin{tabular}{l|lccccc}
  \hline
   & \textbf{Overall} & \multicolumn{4}{c}{By primary target} \\
   & & & Null hyp- & Selected & Predictive &  \\
  Performance & & Estimand & othesis & model & performance & Other \\
  measure & & ($n=64$) & ($n=21$) & ($n=8$) & ($n=3$) & ($n=4$) \\
  \hline
  Convergence               & \textbf{12/85 (14\%)} & 10/61 (16\%) & 1/15 (7\%) & 1/6 (17\%) & 0/2 & 0/1 \\
  Bias                      & \textbf{63/80 (79\%)} & 59/64 (92\%) & 1/9 (11\%) & 0/2 & 2/3 & 1/2 \\
  Empirical SE              & \textbf{31/78 (40\%)} & 31/62 (50\%) & 0/9 & 0/2 & 0/3 & 0/2 \\
  Mean squared error        & \textbf{26/78 (33\%)} & 22/62 (35\%) & 2/9 (22\%) & 0/2 & 1/3 & 1/2 \\
  Model SE                  & \textbf{22/77 (29\%)} & 21/62 (34\%) & 1/9 (11\%) & 0/2 & 0/3 & 0/1 \\
  Type I error              & \textbf{31/95 (33\%)} & 8/62 (13\%) & 18/21 (86\%) & 4/6 & 0/3 & 1/3 \\
  Power                     & \textbf{28/95 (29\%)} & 8/63 (13\%) & 14/20 (17\%) & 4/6 & 0/3 & 2/3 \\
  Coverage                  & \textbf{42/79 (53\%)} & 39/63 (62\%) & 1/9 (11\%) & 0/2 & 1/3 & 1/2 \\
  Conf.\ int.\ length       & \textbf{11/80 (14\%)} & 9/63 (14\%) & 0/10 & 0/2 & 1/3 & 1/2 \\
  \hline
\end{tabular} \\
Note -- denominator changes across performance measures because not all are applicable in all simulation studies
\end{table}

Description and estimation of common performance measures of interest are given in section \ref{s:analysis}. An important point to appreciate in design and analysis is that simulation studies are empirical experiments, meaning performance measures are themselves estimated, and estimates of performance are thus subject to error. This fundamental feature of simulation studies does not seem to be widely appreciated, as previously noted\cite{Ripley87}. The implications are two-fold. First, we should present estimates of uncertainty (quantified as the Monte Carlo standard error; see section \ref{ss:pms}). Second, we need to consider the number of repetitions $\ns$ and how this can be chosen (see section \ref{ss:pms}).


\section{Computational and programming issues in simulation studies} \label{s:coding}

In this section we discuss consideration when coding a simulation study. It is useful to understand what sort of data are involved. There may be up to four classes of dataset, listed and described in table \ref{t:datasets}.
\begin{table}
\caption{The different datasets that may be involved in a simulation study\label{t:datasets}}
\centering
\begin{tabular}{p{2cm} p{10cm}}
  \hline
  Dataset & Description and notes \\
  \hline
  Simulated & A dataset of size $\no$ produced with some element of random-number generation, to which one or more methods are applied to produce some quantity relating to the \textit{target} of the study, such as an estimate of $\theta$. \\
  Estimates$^{\dagger}$ & Dataset containing $\ns$ summaries of information from repetitions (e.g.\ $\hat{\theta}$, $\widehat{\text{SE}}(\hat{\theta})$, indication of hypothesis rejection) for each combination of data-generating mechanism, method and target (e.g.\ each estimand). \\
  States & Dataset of containing $\ns +1$ random-number-generator states: the start state for each simulated dataset and the final state (see section~\ref{ss:rng}). \\
  Performance measures & Dataset containing estimated performance and Monte Carlo standard errors for each data-generating mechanism, method and target. \\
  \hline
\end{tabular} \\
\textdagger or corresponding summaries for non-estimand targets
\end{table}


\subsection{Random numbers: setting seeds and storing states} \label{ss:rng}

All statistical packages capable of Monte Carlo simulation use a pseudo-random-number generator. Each random number is a deterministic function of the current `\textit{state}' of the random-number generator. After a random number is produced, the state changes, ready to produce the next random number. Because the function is deterministic, the state can be set. Typically, the state is set using a `\textit{seed}'. Seeds do not necessarily map 1:1 to states, and provide doors onto the path of possible states. After enough random-number draws (a very large number in software using modern pseudo-random-number generators), the state will eventually repeat: the path is circular.

The `pseudo' element to random-number generators is sometimes characterised as negative. This is perhaps an artefact of the fact that some early algorithms provided very poor imitations of random numbers. However, modern-era algorithms such as the Mersenne Twister do not suffer from these problems and can, for simulation purposes, be regarded as truly random when used correctly. The toss of a coin or roll of a die may be regarded as equally deterministic, albeit the result of a complex set of unknown factors that act in an uncontrollable fashion. These are not denigrated with the term `pseudo-random': in statistical teaching they are often given as the ultimate example of randomness. However, many stage magicians can control the flip of a coin! If a computer pseudo-random number generator is sufficiently unpredictable and passes the various tests for randomness, it is churlish to regard the `pseudo' aspect as a weakness.

There are several \textit{positive} implications of using a deterministic and reproducible process for generating random numbers. First, if the number of repetitions is regarded as insufficient, the simulation study can continue from its end state. Second, and more importantly, if a certain repetition results in some failure such as non-convergence, the starting state for that repetition can be noted and the repetition re-run under that state, enabling better understanding of when the method does not work so that issues leading to non-convergence can be tackled. Finally, the whole simulation study can be independently run by other researchers, giving the potential for exact (rather than approximate) reproduction of results and the scope for additional methods to be included.

Our practical advice for utilising the deterministic nature of random-number generators is simple but strong: 1.~\textit{set the seed at the beginning, once and only once}; 2.~\textit{store the state of the random-number generator often} (ideally once at the beginning of each repetition and once \textit{following} repetition $i=\ns$). This is important; the following chunk of pseudocode demonstrates the concept:\\
\begin{alltt}
SET RandomSeed to \# \\
FOR Repetition 1 to n_sim by 1 \\
\indent STORE Repetition and RandomNumberState in StatesData[Repetition] \\
\indent GENERATE simulated dataset \\
\indent \ldots \\
END FOR
STORE n_sim+1 and RandomNumberState in StatesData[n_sim+1]
\end{alltt}

The reason for this advice is to avoid unintended dependence between simulated datasets. We will illustrate our caution: one undesirable method of knowing the states for $\ns$ repetitions is to set an initial seed and generate a single vector of length $\ns$ by recording the starting state, generating a single random number, recording the new state, and so on. For the simulation itself, the seed for the $i$th repetition is then set to the $i$th element. To clarify the problem, let $\no = \ns = 4$ and let the first simulation step be generation of vector $x$ from a Uniform(0,1) distribution. The first repetition simulates $x_1$ (which changes the random number state four times) and proceeds. The second repetition then simulates $x_2$, which is made up of observations 2 to 4 from repetition $i=1$ and just one new value. Run in Stata 15 (see supplementary material), the resulting draws of $x$ for the four repetitions are:
\begin{align*}
  x_1 = & ~({\color{black!40!white}0.1338766}, {\color{black!70!white}0.1364070}, 0.4512149, \mathbf{0.0210242}) \\
  x_2 = & ~({\color{black!70!white}0.1364070}, 0.4512149, \mathbf{0.0210242}, 0.3508981) \\
  x_3 = & ~(0.4512149, \mathbf{0.0210242}, 0.3508981, {\color{black!70!white}0.9113581}) \\
  x_4 = & ~(\mathbf{0.0210242}, 0.3508981, {\color{black!70!white}0.9113581},  {\color{black!40!white}0.4707521})
\end{align*}
Note that elements with the same shading contain the same values across rows: The fourth element of $x_1$ is the first element of $x_4$ and appears in all repetitions. Only when $i>\no$ is the draw of $x$ actually independent of the first repetition. Such dependency in simulated data can compromise both performance estimates and Monte Carlo SEs and must be avoided. 

\subsubsection{`Stream' random numbers}
It is common for parts of simulation studies -- fractions of all the repetitions, for example -- to be run in parallel on different cores of high-performance computers (which this article will not mention further). If the advice to set the seed once only is followed, the implication for parallelisation is that, while runs for \textit{different} data-generating mechanisms may be parallelised, it is inadvisable to parallelise repetitions \textit{within} a specific data-generating mechanism.\\ 
\begin{tikzpicture}
  \draw[very thin,black!20!white] (.5,0) -- (3.5,0);
  \draw[very thin,black!20!white] (2,-1.5) -- (2,1.5);
  \draw[black!60!white,dashed] (-1,0) circle (1.3cm);
  \draw[->,red!60!black,opacity=.7,very thick] (-1,1.32) arc (90:10:1.32cm);
  \draw[->,teal,opacity=.7,very thick] (-.081,.919) arc (45:-35:1.3cm);
  \node[align=center] at (-1,.3) {Uncontrolled};
  \node[align=center] at (-1,-.3) {parallel runs};
  \draw[black!40!white,dashed] (2,0) circle (1.3cm);
  \draw[->,red!60!black,opacity=.7,very thick] (2,1.3) arc (90:10:1.3cm);
  \draw[->,teal,opacity=.7,very thick] (3.3,0) arc (0:-80:1.3cm);
  \node[align=center] at (2,.3) {Stream};
  \node[align=center] at (2,-.3) {parallel runs};
\end{tikzpicture}

Suppose we wish to parallelise two sets of $\ns/2$ repetitions. Any simulation study will use random numbers (in order) from a section of the circle. Here, each set of repetitions is represented by a clockwise arrow, and uses $80^{\circ}$ of the total $360^{\circ}$ of random numbers available in the full circle (a caricature for illustrative purposes; in practice, a much smaller fraction would be used). The seed dictates the position on the circle at which an arrow begins (and thus ends). The random numbers used up by the first $\ns/2$ repetitions are represented by the red arrow and for the second $\ns$ by the blue arrow. The left circle depicts two chunks run in parallel with two different, arbitrarily-chosen starting seeds. By chance, they may overlap as seen. This would be a cause for concern. The right circle uses separate streams of random numbers. This breaks the circle into quadrants, and setting the same value of a seed within a stream means that the separate chunks will start at the equivalent point on the quadrants and here there is no chance that one stream will enter another. In the absence of streams, repetitions should not be parallelised for the same data-generating mechanism.

In Stata (version 15 or newer), the stream is set with \\
\texttt{. set rngstream \# } \\
prior to setting the seed. In \textsc{Sas}, it is achieved within a\texttt{ data }step with \\
\texttt{. call stream(\#);} \\
In R, this can be achieved with the\texttt{ rstream }package.
Regardless of the package, the same seed must be used within different values of \#.

When a simulation study uses multiple data-generating mechanisms, these may be run in parallel. Because performance is typically estimated separately for different data-generating mechanisms, using the same seeds is less of a problem (and may in fact be advantageous, as described in section~\ref{ss:ran}).

Many programs execute methods involving some stochastic element. Examples include multiple imputation, the bootstrap, the g-computation formula, multistate models and Bayesian methods that use Markov Chain Monte Carlo. Commands to implement these methods involve some random-number generation. It is important to check that such programs do not manipulate the seed. Some packages do have a default seed if not input by the user. If they do set the seed internally, many of the $\ns$ results will be highly correlated, if not identical, and results should not then be trusted. Checking for such behaviour is worthwhile. One simple technique is to display the current state of the random-number generator, twice issue the command, and display the state after each run. If the first and second states are the same then the program probably does not use random numbers. If the first and second states differ but the second and third do not, the seed is being reset by the program.

\subsection{Start small and build up code} \label{ss:build}

As with any coding task it is all-too-easy to obtain misleading results in a simulation study through very minor coding errors; see for example the comments section of \cite{statgeek16bsmi}, where fixing an error in a single line of code completely changed the results. A function may be sloppily written as\texttt{ a-b*c }such that it is unclear if\texttt{ (a-b)*c }or\texttt{ a-(b*c) }was intended; a machine will interpret this code but will not discern the intention.

Errors are often detected when results are unexpected: for example, when bias appears much greater than theory suggests. One design implication is that methods with known properties should be included where possible as a check that these properties are exhibited. One straightforward and intuitive approach for minimising errors is to start small and specific for one repetition, then build and generalise, including plenty of built-in checks.

In a simulation study with $\ns>1$ and several simulated variables, a good starting point is to generate one simulated dataset with large $\no$. If variables are being generated separately then the code for each should be added one by one and the generated data explored to 1) check that the code behaves as expected and 2) ensure the data have the desired characteristics. For example, Stata's \texttt{ rnormal(m,s) }function simulates normal variates with mean\texttt{ m }and standard deviation\texttt{ s}. The usual notation for a normal distribution uses a mean and \textit{variance}. We have seen this syntax trip up several good programmers. By checking the standard deviation of a variable simulated by\texttt{ rnormal() }in a single large simulated dataset, it should be obvious if it does not behave in the expected fashion. The simulation file should be built to include different data-generating mechanisms, methods or estimands, again checking that behaviour is as expected. Using the above example again, if the basic data-generating mechanism used $\text{N}(\mu ,1)$, the issue with specifying standard deviations \textit{vs.}\ variances would not be detected, but it would for data-generating mechanisms with $\sigma^2 \neq 1$. When satisfied with the large dataset being generated, we apply each method.

Once satisfied that one large run is behaving sensibly, it is worth setting the required $\no$ for the simulation study and exploring the simulated datasets produced under a handful of different seeds. When satisfied that the program still behaves sensibly, it may be worth running a few (say tens of) repetitions. If for example convergence problems are anticipated, or bias is expected to be 0, this can be checked informally without the full set of simulations.

After thoroughly checking through and generalising code, the full set of $\ns$ repetitions may be run. However, recall the precaution in section~\ref{ss:rng} to store the states of the random-number generator and the reasons. If failure occurs in repetition $4,120$ of $5,000$, we will want to understand why. In this case, a record of the $4,120$th start state means we can reproduce the problematic dataset quickly.

While the ability to reproduce specific errors is useful, it is also practically helpful to be able to continue even when an error occurs. For this purpose, we direct readers to the\texttt{ capture }command in Stata and the\texttt{ try }command in R. The failed analysis must be recorded as a missing value in the Estimates dataset, together with reasons if possible.

\subsection{Using different software packages for different methods} \label{ss:software}

It is frequently the case that competing methods are implemented in different software packages, and it would be more burdensome to try and code them all in one package than to implement them in different packages. There are two possible solutions. The first is to simulate data separately in the different packages and then use the methods on those data. The second is to simulate data in one package and export simulated data so that different methods are based on the same simulated datasets.

Both approaches are valid in principle but we advocate the latter. First, if data are generated independently for different methods, there will be different (random) Monte Carlo error affecting each repetition. By using the same simulated data for both comparisons, this Monte Carlo error will affect methods' performance in the same way because methods are matched on the same generated data. Second, it is cumbersome to do a job twice, and because different software packages have different quirks, it will not be easy to ensure data really are being generated identically. Third, it is important to understand that our aim is to compare methods, and while the software implementation may be important to evaluate, the way the software package simulates data is not of interest: using a method in practice would involve a software implementation, but not simulating data using that package. Whatever data an analyst was faced with would be the same regardless of the software being used.

Sixty two simulation studies in our review mentioned software. Table \ref{t:software} (in the appendix) describes the specific statistical software mentioned. Seven simulation studies mentioned using more than one statistical package.

\section{Analysis of estimates data} \label{s:analysis}
This section describes estimation for various performance measures along with Monte Carlo SEs. We advocate two preliminaries: checking for missing estimates and plots of the estimates data.

\subsection{Checking the estimates data and preliminaries} \label{ss:prelims}
The number of missing values, e.g.\ of $\hat{\theta}_i$ and $\widehat{\text{SE}}(\hat{\theta}_i)$ (for example due to non-convergence), is the first performance measure to assess. The data produced under repetitions for which missing values were returned should be explored to understand how a method failed (see section \ref{s:coding}) and, ideally, the code made more robust to reduce the frequency of failures.

Missing values in the \textit{estimates} dataset pose a missing data problem regarding the analysis of other performance measures. It seems implausible that values would be missing completely at random\cite{Rubin76}; estimates will usually be missing due to non-convergence so will likely depend on some characteristic/s of a given simulated dataset. When the `method' being evaluated involves an analyst's procedure (as described in section \ref{ss:meth}), for example, the model changes if the first-choice model does not converge, this can reduce or remove missing values from the estimates data (though it changes the nature of the method being evaluated; see section \ref{ss:meth}).

If more than two methods are evaluated, and one always returns an estimate $\hat{\theta}_i$, then missing values for another method may be related to the returned values for the first method. In the presence of a non-trivial proportion of missing estimates data, analysis of further performance measures should be tentative, particularly when comparing methods with different numbers of $\hat{\theta}_i$ missing. `Non-trivial' means any proportion that could meaningfully alter estimated performance. If we are interested in detecting tiny biases, even 1\% may be non-trivial.

Before undertaking a formal analysis of the estimates dataset, it is sensible to undertake some exploratory analysis. 
Plots are often helpful here. For example, \cite{kahan13factor} assessed the performance of a two-stage procedure for the analysis of factorial trials. The procedure was unbiased (both conditionally and unconditionally), yet a histogram of $\hat{\theta}_i$ exhibited a bimodal distribution with modes equally spaced at either side of $\theta$, with almost no values of $\hat{\theta}_i$ close to $\theta$. This may cause concern and would have been missed had the analysis proceeded straight to the estimation of performance.

For simulation studies targeting an estimand, the following plots are often informative:
\begin{enumerate}
  \item A univariate plot of the distribution of $\hat{\theta}_i$ and $\widehat{\text{SE}}(\hat{\theta}_i)$ for each data-generating mechanism, estimand and method, to inspect the distribution and, in particular, to look for outliers.
  \item A bivariate plot of $\widehat{\text{SE}}(\hat{\theta}_i)$ \textit{vs.}\ $\hat{\theta}_i$ for each data-generating mechanism, estimand and method, with the aim of identifying bivariate outliers.
  \item Bivariate plots of $\hat{\theta}_i$ (and possibly $\widehat{\text{SE}}(\hat{\theta}_i)$) for one method \textit{vs.}\ another for each data-generating mechanism and estimand. The purpose here is to look for correlations between methods and any systematic differences. Where more than two methods are compared, a graph of every method \textit{vs.}\ every other or \textit{vs.}\ the comparator can be useful.
  \item Limits-of-agreement for $\hat{\theta}_i$ (and possibly $\widehat{\text{SE}}(\hat{\theta}_i)$) compared with a reference method. That is, a plot of the difference \textit{vs.}\ the mean of each method compared with a comparator.
  \item A plot of confidence intervals fractionally ranked by $|z|$ where $z=(\hat{\theta}_i-\theta)/\text{ModSE}_i$ (as in figure~\ref{f:zipplot}). This is called a zip~plot and is a means of understanding any issues with coverage.
\end{enumerate}
These plots will be demonstrated and interpreted in the worked example (section \ref{s:egsims}).

\subsection{Estimation of performance and Monte Carlo standard errors for some common performance measures} \label{ss:pms}
This section outlines some common performance measures, properties they are designed to assess, how they are estimated and how Monte Carlo standard errors are computed. We suppress the `hat' notation for performance measures, but emphasise that these are estimates.

For interpretation of results, performance measures should usually be considered jointly (one could prefer a method with zero variance by conveniently ignoring bias).

Monte Carlo standard errors quantify simulation uncertainty: they provide an estimate of the SE of (estimated) performance due to using finite $\ns$. The Monte Carlo SE targets the sampling distribution of repeatedly running the same simulation study (with $\ns$ repetitions) under different random-number seeds.

In our review of simulation studies in \textit{Statistics in Medicine} Volume~34, 93 did not mention Monte Carlo SEs for estimated performance. The formulas for computing Monte Carlo SEs given in table \ref{t:pms} with description and comments in the text. For empirical SE, relative \% increase in precision, and relative error, the Monte Carlo SE formulas assume normally distributed $\hat{\theta}$; for non-normal $\hat{\theta}$, robust SEs exist -- see \cite{white10}.

\begin{table} \centering
\caption{Performance measures: definitions, estimates and Monte Carlo standard errors \label{t:pms}}
\begin{tabular}{llll}\hline
Performance & & & \\
measure & Definition & Estimate & Monte Carlo SE of estimate\\
\hline
 &&&\\
Bias
  & $\text{E}[\hat\theta]-\theta$
  & $\dfrac{1}{\ns}\sum\limits_{i=1}^{\ns} \hat{\theta}_i-\theta$
  & $\sqrt{\frac{1}{\ns(\ns -1)}\sum\limits_{i=1}^{\ns}(\hat{\theta}_i-\bar\theta)^2}$\\
 &&&\\
EmpSE
  & $\sqrt{\text{Var}(\hat\theta)}$
  & $\sqrt{\dfrac{1}{\ns -1}\sum\limits_{i=1}^{\ns}(\hat{\theta}_i-\bar{\theta})^2}$
  & $\frac{\widehat{\text{EmpSE}}}{\sqrt{2(\ns-1)}}$\\
 &&&\\
\parbox{8em}{Relative \% increase in precision (B \textit{vs}.\ A) *}
  & $100\left(\frac{\text{Var}(\hat{\theta}_A)}{\text{Var}(\hat{\theta}_B)}-1\right)$
  & $100\left(\left(\frac{\widehat\text{EmpSE}_A}{\widehat\text{EmpSE}_B}\right)^2-1\right)$
  & $200\left(\frac{\widehat\text{EmpSE}_A}{\widehat\text{EmpSE}_B}\right)^2 \sqrt{\frac{1-\text{Corr}(\hat{\theta}_A, \hat{\theta}_B)^2}{\ns-1}}$\\
 &&&\\
MSE
  & $\text{E}[(\hat{\theta}-\theta)^2]$
  & $ \dfrac{1}{\ns}\sum\limits_{i=1}^{\ns}(\hat{\theta}_i-\theta)^2$
  & $\sqrt{\frac{\sum\limits_{i=1}^{\ns}\left[(\hat{\theta}_i-\theta)^2-\widehat{\text{MSE}}\right]^2}{\ns(\ns-1)}}$ \\
 &&&\\
Average ModSE *
  & $\sqrt{\text{E}[\widehat{\text{Var}}(\hat{\theta})]}$
  & $\sqrt{\frac{1}{\ns}\sum\limits_{i=1}^{\ns}\widehat{\text{Var}}(\hat{\theta}_i)}$
  & $\sqrt{\frac{\widehat{\text{Var}}[\widehat{\text{Var}}(\hat{\theta})]}{4\ns\times \widehat{\text{ModSE}}^2}} ~~\dagger$ \\
 &&&\\
\parbox{8em}{Relative \% error in ModSE *}
  & $100\left(\frac{\text{ModSE}}{\text{EmpSE}}-1\right) $
  & $100\left(\frac{\widehat{\text{ModSE}}}{\widehat\text{EmpSE}}-1\right) $
  & $100\left(\frac{\widehat\text{ModSE}}{\widehat\text{EmpSE}}\right) \sqrt{\frac{\widehat{\text{Var}}[\widehat{\text{Var}}(\hat{\theta})]}{4\ns\times \widehat{\text{ModSE}}^4} + \frac{1}{2(n-1)}} ~~\dagger$\\
 &&&\\
Coverage
  & $\text{Pr}(\hat{\theta}_{\text{low}}\leq \theta \leq \hat{\theta}_{\text{upp}})$
  & $\frac{1}{\ns}\sum\limits_{i=1}^{\ns} 1(\hat{\theta}_{\text{low},i}\leq \theta \leq \hat{\theta}_{\text{upp},i})$
  & $\sqrt{\frac{\widehat\text{Cover.} \times (1-\widehat\text{Cover.})}{\ns}}$\\
 &&&\\
\parbox{8em}{Bias-eliminated coverage}
  & $\text{Pr}(\hat{\theta}_{\text{low}}\leq \bar{\theta} \leq \hat{\theta}_{\text{upp}})$
  & $\frac{1}{\ns}\sum\limits_{i=1}^{\ns} 1(\hat{\theta}_{\text{low},i}\leq \bar{\theta} \leq \hat{\theta}_{\text{upp},i})$
  & $\sqrt{\frac{\widehat\text{B-E~Cover.} \times (1-\widehat\text{B-E~Cover.})}{\ns}}$\\
 &&&\\
\parbox{8em}{Rejection \% (power or type~I error)}
  & $\text{Pr}(p_i\leq\alpha) $
  & $\frac{1}{\ns}\sum\limits_{i=1}^{\ns} 1(p_i\leq\alpha) $
  & $\sqrt{\frac{\widehat\text{Power} \times (1-\widehat\text{Power})}{\ns}}$\\
 &&&\\
\hline
\end{tabular} \\
\parbox{\textwidth}{*Monte Carlo SEs are approximate for \textit{Relative \% increase in precision}, \textit{Average ModSE} and \textit{Relative \% error in ModSE}. \\
$\dagger \widehat{\text{Var}}[\widehat{\text{Var}}(\hat{\theta})]
=
\dfrac{1}{\ns-1}\sum\limits_{i=1}^{\ns} \left\{\widehat{\text{Var}}(\hat{\theta}_i) - \dfrac{1}{\ns}\sum\limits_{j=1}^{\ns}\widehat{\text{Var}}(\hat{\theta}_j) \right\}^2$.
}
\end{table}

Bias is frequently of central interest, and quantifies whether a method targets $\theta$ on average. Frequentist theory holds unbiasedness to be a key property.

The mean of $\hat{\theta}_i$, $\bar{\theta}$, is often reported instead. This is estimated in the same way but without subtracting the constant $\theta$, and so has the same Monte Carlo SE. It is sometimes preferable to report the relative bias, rather than absolute. If different values of $\theta$ are used for different data-generating mechanisms then relative bias permits a more straightforward comparison across values. However, relative bias can be used only for $\lvert \theta \rvert > 0$. The absence of bias is one property of an estimator; while it is often of central interest, we may sometimes accept small biases because of other good properties.

The empirical SE is a measure of the precision or efficiency of the estimator of $\theta$. It depends only on $\hat{\theta}_i$ and does not require knowledge of $\theta$. The empirical SE estimates the long-run standard deviation of $\hat{\theta}_i$ over the $n_\text{sim}$ repetitions. Several other designations are in common use; in our review, the terms used included `empirical standard deviation', `Monte Carlo standard deviation', `observed SE', and `sampling SE'.

The empirical standard error can be hard to interpret for a single method (unless compared to a lower bound), and the relative precision is often of interest when comparing methods.

Note that, if either method is biased, relative precision should be interpreted with caution because an estimator that is biased towards the null can have small empirical~SE as a result of the bias: $\hat{\theta_i}/2$ has smaller empirical SE than $\hat{\theta_i}$.

\begin{figure} \centering
  \caption{The impacts of bias and empirical SE on root MSE and coverage of nominal 95\% confidence intervals, compared for three methods: Method~A is unbiased but imprecise; Method~B is biased (independent of $\no$) and more precise; Method~C is biased (with bias $\propto \sqrt{1/\no}$) and the same precision as method~B. The comparison of root~MSE and coverage depends on the choice of $\no$; the constant bias of method~B dominates its increasingly poor MSE and coverage as $\no$ increases \label{f:mse}}
\begin{tikzpicture}[baseline]
\begin{axis}[
    samples=100,
    domain=12:90,
    height=4cm,
    width=5cm,
    xlabel=$\no$,
    title=Bias,
    ymin=-.1,ymax=3.1,
    xmin=10,xmax=90,
    xtick={10,30,50,70,90},
    ymajorgrids,
    grid style={draw=black!20},
    tick style={draw=black!20},
    axis line style={draw=none}
  ]
  \addplot[v6,very thick] plot(\x,{0});
  \addplot[v3,very thick] plot(\x,{1});
  \addplot[v10,very thick] plot(\x,{sqrt(40/x)});
  \node[v6,align=center] at(130,35) {A};
  \node[v3,align=center] at(130,85) {B};
  \node[v10!60!black,align=center] at(130,170) {C};
\end{axis}
\end{tikzpicture}
\begin{tikzpicture}[baseline]
\begin{axis}[
    samples=100,
    domain=12:90,
    height=4cm,
    width=5cm,
    xlabel=$\no$,
    title=Empirical SE,
    ymin=-.1,ymax=3.1,
    xmin=10,xmax=90,
    xtick={10,30,50,70,90},
    ymajorgrids,
    grid style={draw=black!20},
    tick style={draw=black!20},
    axis line style={draw=none}
  ]
  \addplot[v6,very thick] plot(\x,{sqrt((10^2)/x});
  \addplot[v3,very thick] plot(\x,{.05+sqrt((6^2)/x});
  \addplot[v10,very thick,opacity=.6] plot(\x,{sqrt((6^2)/x});
  \node[v6,align=center] at(500,170) {A};
  \node[v10!50!v3,align=center] at(250,55) {B \& C};
\end{axis}
\end{tikzpicture}
\begin{tikzpicture}[baseline]
\begin{axis}[
    samples=100,
    domain=12:90,
    height=4cm,
    width=5cm,
    xlabel=$\no$,
    title=$\rightarrow$ Root MSE,
    ymin=-.1,ymax=3.1,
    xmin=10,xmax=90,
    xtick={10,30,50,70,90},
    ymajorgrids,
    grid style={draw=black!20},
    tick style={draw=black!20},
    axis line style={draw=none}
  ]
  \addplot[v6,very thick] plot(\x,{sqrt((10^2)/x});
  \addplot[v3,very thick] plot(\x,{sqrt((6^2)/x + 1)});
  \addplot[v10,very thick] plot(\x,{sqrt((6^2)/x + (40/x))});
  \node[v6,align=center] at(150,250) {A};
  \node[v3,align=center] at(70,140) {B};
  \node[v10!60!black,align=center] at(500,75) {C};
\end{axis}
\end{tikzpicture}
\begin{tikzpicture}[baseline]
\begin{axis}[
    samples=100,
    domain=2:98,
    height=4cm,
    width=5cm,
    xlabel=$\no$,
    title=$\rightarrow$ Coverage \%$^\star$,
    ymin=38,ymax=102,
    xmin=2,xmax=98,
    ytick={40,60,80,100},
    xtick={10,30,50,70,90},
    ymajorgrids,
    grid style={draw=black!20},
    tick style={draw=black!20},
    axis line style={draw=none}
  ]
  \addplot[v6,very thick] plot(\x,{
    100*(normcdf(1.96) - normcdf(-1.96))
  });
  \addplot[v3,very thick] plot(\x,{
    100*(normcdf((1/(5*x^(-.5)))+1.96) - normcdf((1/(5*x^(-.5)))-1.96))
  });
  \addplot[v10,very thick] plot(\x,{
    100*(normcdf(1+1.96) - normcdf(1-1.96))
  });
  \node[v6,align=center] at(750,610) {A};
  \node[v3,align=center] at(350,300) {B};
  \node[v10!60!black,align=center] at(750,360) {C};
\end{axis}
\end{tikzpicture}\\
$^\star$Coverage calculated for normal-based confidence intervals of correct width so that bias is the source of undercoverage
\end{figure}
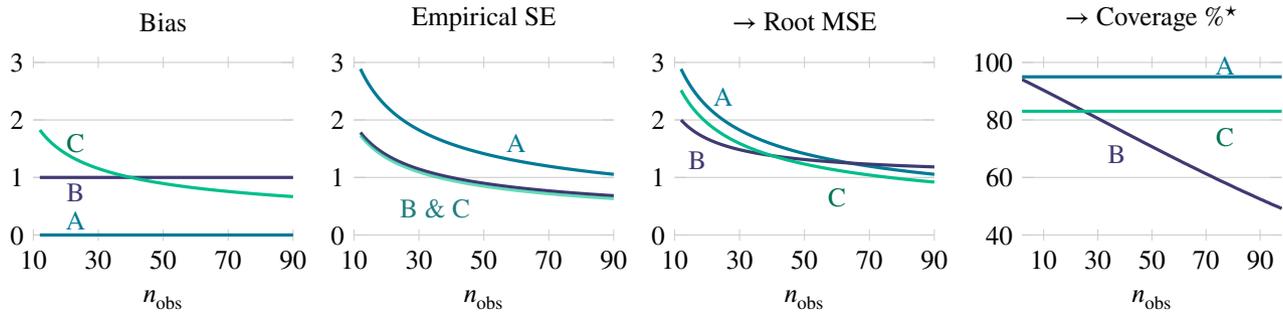

A related measure, which also takes the true value of $\theta$ into account, is the mean squared error (MSE). The MSE is the sum of the squared bias and variance of $\hat{\theta}$. This appears a natural way to integrate both measures into one summary performance measure (low variance is penalised for bias), but we caution that, for method comparisons, the relative influence of bias and of variance on the MSE tends to vary with $\no$ (except when all methods are unbiased), making generalisation of results difficult. This issue is illustrated in the first three panels of figure \ref{f:mse}, which depict the bias, empirical standard error and root MSE (favoured here because it is on the same scale as Empirical SE) for three hypothetical methods. Method~A is unbiased but imprecise (and so root MSE is simply the empirical SE); method~B is biased, with bias constant over $\no$, but more precise (as is often the case with biased methods, see for example \cite{White09}); and method~C, which is biased in a different way (bias $\propto \sqrt{1/\no}$) and with precision the same as method~B. For $\no <60$, root MSE is lower for method~B than A, but for $\no >60$, root MSE is lower for method~A. The lesson is that comparisons of (root) MSE are more sensitive to choice of $\no$ than comparisons of bias or empirical SE alone. MSE is nonetheless an important performance measure -- particularly when the aims of a method relate to prediction rather than estimation -- but the implication is that, when MSE is a performance measure, data-generating mechanisms should include a range of values of $\no$. \bigskip

\noindent We term the root-mean of the squared model SEs the `average model~SE'. The aim of the model~SE is that $\text{E}(\text{ModSE}^2)=\text{EmpSE}^2$. The model~SE explicitly targets the empirical~SE. If it systematically misses, this represents a bias in the estimation of model~SE. The relative error in average model~SE is then an informative performance measure (some prefer the ratio of average model~SE to empirical~SE).

Coverage of confidence intervals is a key property for the long-run frequentist behaviour of an estimator. It is defined as the probability that a confidence interval contains $\theta$.

Note that Neyman's original description of confidence intervals defined the property of \textit{randomisation validity} as exactly $100(1-\alpha)\%$ of intervals containing $\theta$ (see \cite{neyman34,Meng94,Rubin96}). \textit{Confidence validity} is the property that the true percentage is at least $100(1-\alpha)$. This latter definition is less well known than the former, with the result that over- and under-coverage are sometimes regarded as similarly bad\cite{morris12}. Of course, randomisation validity would usually be preferred over confidence validity because it implies shorter intervals -- but this is not always the case! There are examples of procedures that return both shorter intervals and higher coverage (see for example \cite{Meng94,Rubin96}).

Under-coverage is to be expected if, for example, i) $\text{Bias}\neq 0$, ii) $\text{ModSE}<\text{EmpSE}$, iii) the distribution of $\hat{\theta}$ is not normal and intervals have been constructed assuming normality, or iv) $\widehat{\text{Var}}(\hat{\theta}_i)$ is too variable. Over-coverage tends to occur as a result of $\text{ModSE}^2>\text{EmpSE}^2$. This may occur either in the absence or presence of issues (i) and (iii).

Under-coverage due to bias will tend to deteriorate as $\no$ increases (unless bias reduces at or faster than a rate of $\sqrt{1/\no}$~). Intuitively, as $\no$ increases, confidence intervals zero-in on the wrong value. The situation is illustrated in the fourth panel of figure~\ref{f:mse} (with some loss of generality). Coverage was calculated for normal-based confidence intervals with correct interval width such that bias is the only source of under-coverage. Method~B, for which bias is independent of $\no$, has deteriorating coverage as $\no$ increases. Method~C, for which $\text{bias}\propto \sqrt{1/\no}$, has constant under-coverage. As for MSE, bias dominates as $\no$ increases. The implication is that, if coverage is being assessed in the presence of bias, the data-generating mechanisms should include a range of values of $\no$.

As noted previously, there are two sources of poor coverage: bias leads to under-coverage, while incorrect interval width (for example because $\text{average~ModSE} \neq \text{EmpSE}$) may produce under- or over-coverage. We propose a decomposition of poor coverage with a new performance measure: `bias-eliminated coverage'. By studying confidence interval coverage for $\bar{\theta}$ rather than for $\theta$, the bias of a method is eliminated from the calculation of coverage. We emphasise that bias-eliminated coverage should not be regarded as a performance measure in its own right. It is to be used for understanding how coverage performance is influenced by bias \textit{vs.}\ width of confidence intervals. It is obvious that, for the methods in figure~\ref{f:mse}, bias-eliminated coverage will be equal for all methods. 

Rejection rates -- power and type~I error -- are often of principal interest in simulation studies that target a null hypothesis, and power is of particular interest when competing designs are being compared by simulation. Assume we have \textit{p}-values $p_i$ in the estimates data and are considering nominal significance level $\alpha$. The \textit{p}-values may be derived from a Wald statistic $\frac{\hat{\theta}_i}{\widehat{\text{SE}}(\hat{\theta}_i)}$  or output directly, for example by a likelihood-ratio test. An appropriate test would reject a proportion $\alpha$ of the $\ns$ repetitions when the null is true and as often as possible when it is false. The obvious warning is that if the test does not control type~I error at level $\alpha$, power should be interpreted with caution. \bigskip

It is sometimes (not always) of interest to estimate \textit{conditional} performance. This is particularly true for simulation studies that aim to evaluate alternative study designs, for example where design decisions are made based on the early data. Two simulation studies in our review sample explored two-stage procedures in randomised trials, where the estimand is selected after the first stage: the estimand was the treatment effect in a selected subpopulation \cite{kimani15} or the effect of a selected treatment \cite{carreras15}. In both cases, estimators were designed to be conditionally unbiased. Kimani, Todd \& Stallard reported bias conditional on each possible selection of estimand \cite{kimani15}, while Carreras, Gutjahr \& Brannath considered the bias averaged across estimands \cite{carreras15}. The former method is stricter and arguably more appropriate since, having selected an estimand, the observer is not interested in the other case\cite{casi16fisher}.

Now consider the following form of conditional performance: $\no = 30$ observations are simulated from $y \sim \text{N}(\mu,\sigma^2)$, with $\mu=0$, $\sigma^2=1$. For each repetition, 95\% confidence intervals for $\mu$ are constructed using the $t$-distribution. The process is repeated $\ns = 30,000$ times, and we study the coverage, 1)~for all repetitions, and 2)~according to tertiles of the model~SE. The results, given in table \ref{t:cond}, show that coverage is below 95\% for the lowest third of standard errors, above 95\% for the highest third, and slightly above for the middle third. Poor conditional performance in this sense should not cause concern. Further, it is unhelpful for an analyst faced with a dataset: one would not in practice know in which tertile of possible model~SEs a particular model~SE lies.

\begin{table}
  \caption{Coverage conditional on size of ModSE\label{t:cond}}
  \centering
  \begin{tabular}{lrr}
    \hline
    Approach & $\ns$ analysed & Coverage (Monte Carlo \textsc{SE}) \\
    \hline
    All observations & 30,000 & 95.0\% \footnotesize{(0.1\%)} \\
    Conditional: ModSE in highest third & 10,000 & 98.0\% \footnotesize{(0.1\%)} \\
    Conditional: ModSE in middle third & 10,000 & 95.5\% \footnotesize{(0.2\%)} \\
    Conditional: ModSE in lowest third & 10,000 & 91.5\% \footnotesize{(0.3\%)} \\
    \hline
  \end{tabular}
\end{table}

Estimating performance conditional on \textit{true} (rather than sample-estimated) parameters that vary across data-generating mechanisms is where methods \textit{should} be expected to provide good performance, and we do not recommend averaging over these, as is done informally in \cite{gutman13}. \bigskip


\noindent We have described the most commonly reported and generally applicable performance measures, particularly when a simulation study targets an estimand. There are others that are sometimes used (such as the proportion of times the correct dose is selected by dose--finding designs) and others that we have not yet thought of.

\subsection{Sample size for simulation studies} \label{ss:sampsi}
In choosing $\ns$, the central issue is Monte Carlo error: key performance measures need to be estimated to an acceptable degree of precision.

The values of $\ns$ reported in our review are shown in figure \ref{f:dgms}, panel~d. Four simulation studies did not report $\ns$. Common sample sizes are $\ns = 500$ and $\ns = 1,000$, as previously reported by Burton \textit{et al.}\cite{Burton06}. Of the 87 studies reporting $\ns$, four provided any justification of the choice. These were:
\begin{itemize}
  \item `\textit{To evaluate the asymptotic biases}'~\cite{taguri15}
  \item `\textit{errors can be reduced by the large number of simulation replicates}'~\cite{li15}
  \item `\textit{number was determined mainly to keep computing time within a reasonable limit. A reviewer pointed out that, as an additional justification, by using 10,000 meta-analyses the standard error of an estimated percentage (e.g., for the empirical coverage) is guaranteed to be smaller than 0.5.}'~\cite{kuss15}
  \item Most positively, Marozzi gave an explicit derivation of Monte Carlo SE~\cite{marozzi15}
\end{itemize}

Clearly this is a sub-optimal state of affairs. For some more concrete justifications, see the worked illustrative example in section \ref{s:egsims}, Keogh and Morris\cite{keogh18}, or Morris \textit{et al.}\cite{morris17ma12}

There exist situations where only one repetition is necessary, particularly when investigating large-sample bias; see for example \cite{white97}. Here, the aim was to demonstrate large-sample bias of an estimator and the single estimate of $\hat{\theta}$ was many model standard errors from its true value.

Where the key performance measure is coverage, $\ns$ can be defined as follows. The Monte Carlo SE of coverage is given in section \ref{ss:pms}.
Plugging in the expected coverage (for example 95\%) and rearranging, we get
\begin{align} \label{eq:ecov}
  \ns = \frac{\text{E(Coverage)}\times(1-\text{E(Coverage)})}{(\text{Monte Carlo SE}_\text{req})^2}
\end{align}
with a similar expression if $\ns$ is to be determined based on power. For example, if the SE required for a coverage of 95\% is 0.5\%,
  $$\ns = \frac{95 \times 5}{0.5^2} = 1,900 \text{~repetitions.}$$
Coverage is estimated from $\ns$ binary summaries of the repetitions, so the worst-case SE occurs when coverage is 50\%. In this scenario, to keep the required Monte Carlo SE below 0.5\% , (\ref{eq:ecov}) says that $\ns = 10,000$ repetitions will achieve this Monte Carlo SE.

A convenient feature of simulation studies is that the Monte Carlo SE can be assessed and $\ns$ increased much more cheaply than with other empirical studies. The cost is computational time. To continue, rather than start again, it is important to have a record the end state of the random-number generator (which can be used as the seed if further repetitions are added) or to use a different stream.

\subsection{Remarks on analysis}\label{ss:ran}
We have emphasised repeatedly that simulation studies are empirical experiments. In many biomedical experiments, `controls' are used as a benchmark and the estimated effects of other conditions are estimated as a contrast \textit{vs.}\ control. However, simulation studies often benefit from having a known `truth', meaning that the contrast \textit{vs.}\ a control is not often of interest (hence the term `comparator' in section \ref{ss:meth}). That is, bias need not be estimated as the \textit{difference} between $\bar{\theta}-\theta$ for method A and $\bar{\theta}$ for the control; rather the bias for a method stands alone, being computed against $\theta$, the comparator of interest. There are benchmarks for other performance measures as well, such as coverage (the nominal \%) and precision (the Cram\'er--Rao lower bound\cite{cramer46,rao45}).

In some cases, the true value of $\theta$ is unknown: it may not appear in the data-generating mechanism. If performance measures involving $\theta$ are not of interest, this poses no problem. Otherwise, one solution is to \textit{estimate} $\theta$ by simulation. Williamson \textit{et al}.\ simulated data from a logistic model, but $\theta$ was not the conditional odds ratio used to generate data; $\theta$ was the marginal odds ratio, risk ratio and risk difference\cite{williamson14}. They thus estimated $\theta$ for each of these estimands from a large simulated dataset.

In our review of Volume~34, of 74 studies that included some $\theta$, nine estimated it, 57 used a known $\theta$ and 8 were unclear. 
Estimating $\theta$ is in our view a sensible and pragmatic approach. However, such an approach must simulate a dataset so large that it is fair to assume that the variance of `$\theta$' is negligible, particularly compared to that of $\bar{\theta}$, and ensure that the states of the random-number generators used in the simulation study do not overlap with the states used for the purpose of estimating $\theta$. In practice, the way to do this is either to use a separate stream for the random numbers, or to run the $\theta$-estimation simulation immediately before the main run.
\bigskip

The estimation of performance measures in section \ref{ss:pms} is described for estimating performance once per data-generating-mechanism. This is most suited to simulation studies with few data-generating mechanisms, but many simulation studies are considerably more complex. In such cases, it is natural to fit a model (termed `meta-model' by Skrondal\cite{skrondal00}) for performance in terms of the data-generating mechanisms.

An advantage of modeling performance across data-generating mechanisms is that we are able to match repetitions. This reduces the Monte Carlo SE for the comparison of methods. For example, suppose that we have two data-generating mechanisms with $\theta$ equal to 1 and 2. We could use the same starting seed so that results are correlated within $i$.

\section{Reporting} \label{s:report}

\subsection{The `methods' section}
The rationale for the ordering of elements in \textsc{ademp} is that this is usually the appropriate order to report them in a methods section. If the simulation study has been planned and written out before it is executed, the methods section is largely written. This is a particularly helpful ordering for other researchers who might wish to replicate the study. Details should be included to allow reproduction as far as possible, such as the value of $\ns$ and how this was decided on, dependence among simulated datasets. Another important element to report is a justification of the chosen targets for particular applied contexts.

\subsection{Presentation of results}
Some simulation studies can be very small, for example exploring one or two performance measures under a single data-generating mechanism. These can be reported in text (as in He \textit{et al.} \cite{he15}). In other cases, there are enough results that it becomes necessary to report them in tabular or graphical form. For any tabulation or plot of results, there are four potential dimensions: data generating mechanisms, methods, estimands and performance measures. This section provides some considerations for presenting these results.

In tabular displays, it is common to divide rows according to data-generating mechanisms and methods as columns (as in Chen, \textit{et al.}\cite{chen15}), though if there are more methods than data-generating mechanisms it is better to swap these (as in Hsu, Taylor and Hu \cite{hsu15}). Performance measures and estimands may vary across columns or across rows depending on what makes the table easier to digest (see for example Alonso \textit{et al.}\ \cite{alonso15}).

There are two key considerations in the design of tables. The first is how to place the important comparisons side-by-side. The most important comparisons will typically be of methods, so bias (for example) for different methods should be arranged in adjacent rows or columns.

The second consideration regards presentation of Monte Carlo SEs, and this tends to confound the first. By presenting them next to performance results, for example in parentheses, the table becomes cluttered and hard to digest, obscuring interesting comparisons. For this reason, some authors will report the maximum Monte Carle SE in the caption of tables (for example \cite{seaman12,White09}). Results should not be presented to a greater accuracy than is justified by the Monte Carlo SE (e.g.\ 3dp for coverage). In our review of Volume~34, seven articles presented Monte Carlo SEs for estimated performance: three in the text, two in a table, one in a graph, and one in a float caption.

The primary advantage of graphical displays of performance is that it is easier to quickly spot patterns, particularly over dimensions that are not compared side-by-side. A second advantage is that it becomes possible to present raw data estimates (for example the $\hat{\theta}_i$) as well as performance results summarising them (see for example figure 3 of \cite{lambert15}). In our experience, these plots are popular and intuitive ways to summarise the $\hat{\theta}_i$ and model SE's. Another example of a plot of estimates data is a histogram given in Kahan\cite{kahan13factor} (this was particularly important as $\text{Bias}\simeq 0$, but almost no $\hat{\theta}_i$ was close to $\theta$). Even if plots of estimates are not planned to be included in publications we urge their use in exploration of simulation results.

The main disadvantage of graphical displays of results is that plots can be less space-efficient than tables, it is not possible to read the exact numbers, and separate plots will frequently be required for different performance measures.

Compared with tables, it is easier for plots of performance results to accommodate display of Monte Carlo SE's directly, and this should be done, for example as 95\% confidence intervals. The considerations about design of plots to facilitate the most relevant comparisons apply as with tables. Methods often have names that are hard to arrange side by side in a legible manner; it is usually preferable to arrange methods in horizontal rows and performance measures across columns.

As noted previously, full factorial designs can pose problems for presentation of results. One option for presentation is to present data assuming no interaction unless one is obviously present. An alternative approach taken by R\"{u}cker and Schwarzer is to present all results of a full factorial simulation study with $4 \times 4 \times 4 \times 4 \times 3 = 768$ data-generating mechanisms, and comparison of six methods\cite{rucker14}. Their proposal is a `nested-loop plot', which loops through nested factors -- usually data-generating mechanisms -- for an estimand, and plots results for different methods on top of each other\cite{rucker14}. This is a useful graphic but will not suit all designs (and makes depiction of Monte Carlo SE difficult). 

There is no one correct way to present results, but we encourage careful thought to facilitate readability and clarity, considering the comparisons that need to be made by readers.

\section{Worked illustrative example} \label{s:egsims}
To make clear the ideas described in this article and demonstrate how they may be put into practice, we conduct one example simulation study. We hope that the aims and methods are simple enough to be understood by all readers. Further, the files required to run the simulation in Stata are available at https://github.com/tpmorris/simtutorial (with the addition of code for other software planned).

\subsection{Design of example} \label{ss:egdesign}
The example is a comparison of three different methods for estimating the hazard ratio in a randomised trial with a survival outcome.

\begin{figure}
  \centering
  \caption{Visualisation of the true hazard rate over follow-up time in the two data-generating mechanisms. Black (flat) lines are for the first data-generating mechanism where $\gamma=1$; red curves are for the second, where $\gamma=1.5$ \label{f:dgmviz}}
  \begin{tikzpicture}[baseline]
    \begin{axis}[
        samples=200,
        domain=0:5,
        height=.35\textwidth,
        width=.6\textwidth,
        restrict y to domain=0:0.4,
        xlabel=Time $t$ (years), ylabel=$h(t)$ ,
        ymin=0,ymax=.4,
        ymajorgrids,
        grid style={draw=black!10},
        tick style={draw=black!10},
        axis line style={draw=none},
      ]
      \addplot[black,thick] plot(\x,{.1});
      \addplot[black,dashed,thick] plot(\x,{.1*exp(-.5)});
      \addplot[red!80!black,thick] plot(\x,{.1*1.5*(x^.5)});
      \addplot[red!80!black,dashed,thick] plot(\x,{.1*1.5*(x^.5)*exp(-.5)});
      \node[align=center] at(210,280) {$\gamma=1.5, h_i(t)|X=0$};
      \node[align=center] at(380,220) {$\gamma=1.5, h_i(t)|X=1$};
      \node[align=center] at(380,120) {$\gamma=1, h_i(t)|X=0$};
      \node[align=center] at(380,40) {$\gamma=1, h_i(t)|X=1$};
    \end{axis}
  \end{tikzpicture}
\end{figure}
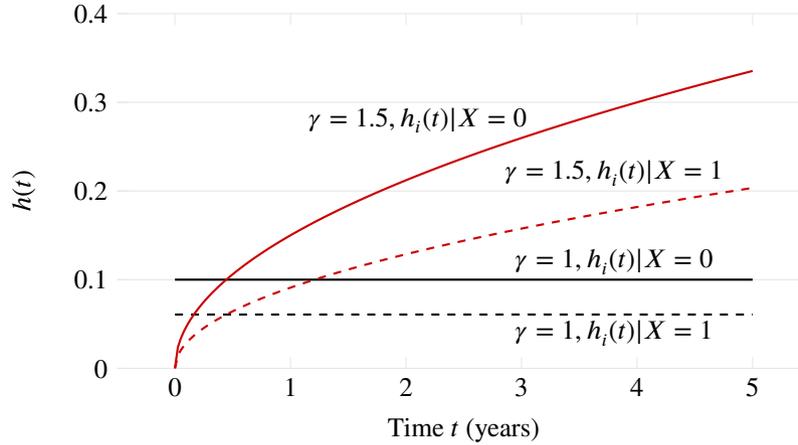

Consider the proportional hazards model, where we have the hazard rate (event rate at time~$t$ conditional on survival until at least time~$t$) for the $i$th patient
\begin{align} \label{eq:ph}
  h_i(t) =&~ h_0(t)\text{exp}(X_i\theta),
\end{align}
with $h_0(t)$ the baseline hazard function, $X_i$ a binary treatment indicator variable coded 0 for control and 1 for the research arm, and $\theta$ the log hazard ratio for the effect of treatment. There are various ways to estimate this hazard ratio, with common approaches being the Cox model, and standard parametric survival models, such as the exponential and Weibull. The parametric approaches make assumptions about the form of the baseline hazard function $h_0(t)$ whereas the Cox model makes no such assumption. We now describe a simulation study to evaluate the three methods in this simple setting. \medskip \\
\noindent\textbf{Aims:} To evaluate the impacts 1) of misspecifying the baseline hazard function on the estimate of the treatment effect $\theta$; 2) of fitting too complex a model when an exponential is sufficient; and 3) of avoiding the issue by using a semiparametric model. \medskip \\
\noindent\textbf{Data-generating mechanisms:} We consider two data-generating mechanisms. For both, data are simulated on $\no = 500$ patients, representing a possible phase III trial with survival outcome. Let $X_i \in (0, 1)$ be an indicator denoting assignment to treatment, where assignment is generated using $X_i \sim \text{Bern}(0.5)$ -- simple randomisation with an equal allocation ratio. We simulate survival times from the model in equation \ref{eq:ph}, assuming that $\theta = -0.5$, corresponding to a hazard ratio of 0.607 (3dp). We let $h_0(t)=\lambda\gamma t^{\gamma-1}$. The two data-generating mechanisms differ only in the values of $\gamma$: \\
\begin{tabular}{lll}
  1: & $\lambda = 0.1, \gamma = 1$   & $\leftarrow$ both an exponential and a Weibull model \\
  2: & $\lambda = 0.1, \gamma = 1.5$ & $\leftarrow$ a Weibull but not an exponential model
\end{tabular} \smallskip \\
A plot of the hazard rate $h_i(t)$ for the two data-generating mechanisms is given in figure \ref{f:dgmviz}.

Data are simulated using Stata 15 using the 64-bit Mersenne twister for random number generation. The input seed is `72789'. \medskip \\
\noindent\textbf{Estimands:} Our estimand $\theta$ is the log-hazard ratio for $X=1$ \textit{vs}.\ $X=0$, which would represent a treatment effect in a randomised trial. \medskip \\
\noindent\textbf{Methods:} Each simulated dataset is analysed in three ways, using:
\begin{list}{.}{\leftmargin=1em\labelwidth=1em\topsep=0em\parsep=0em\itemsep=0em}
  \item[1.] An exponential proportional-hazards model
  \item[2.] A Weibull proportional-hazards model
  \item[3.] A Cox proportional-hazards model
\end{list}
Note that the exponential model is correctly specified for the first data-generating mechanism but misspecified for the second; the Weibull model is correctly specified for both mechanisms; and the Cox model does not make any assumption about the baseline hazard so is not misspecified for either mechanism. \medskip \\
\noindent\textbf{Performance measures:} We will assess convergence, bias, coverage, empirical and model-based standard errors for $\hat{\theta}$.

Bias is our key performance measure of interest, and we will assume that $\text{SD}(\hat{\theta})\leq0.2$, meaning that $\text{Var}(\hat{\theta})\leq 0.04$. (A conservative estimate based on an initial small simulation run.) We decide that we require Monte Carlo SE of bias to be lower than 0.005. Given that $$\text{Monte Carlo SE(Bias)} = \sqrt{\text{Var}(\hat{\theta})/\ns},$$ this implies that we need 1,600 repetitions. If coverage of all methods is 95\%, the implication of using $\ns = 1,600$ is $$\text{Monte Carlo SE(Coverage)}=\sqrt{\frac{95 \times 5}{1,600}} = 0.54.$$ With 50\% coverage, the Monte Carlo SE is maximised at 1.25. We consider this satisfactory and so proceed with $\ns = 1,600$ (to be revised if, for example, $\text{SD}(\hat{\theta})>0.2$).

\subsection{Exploration and visualisation of results} \label{ss:egexplore}
The first result to note is that there were no missing $\hat{\theta}_i$ or $\widehat{\text{SE}}(\hat{\theta}_i)$, and `separation' was not an issue.

\begin{figure} \caption{Plot of the 1,600 $\hat{\theta}_i$ (left panels) and $\widehat{\text{SE}}(\hat{\theta})_i$ (right panels) by data-generating mechanisms, for the three analysis methods. The vertical axis is repetition number, to provide some separation between points. The yellow pipes are sample means.\label{f:thetai}}
  \centering
  \includegraphics[width=.8\textwidth]{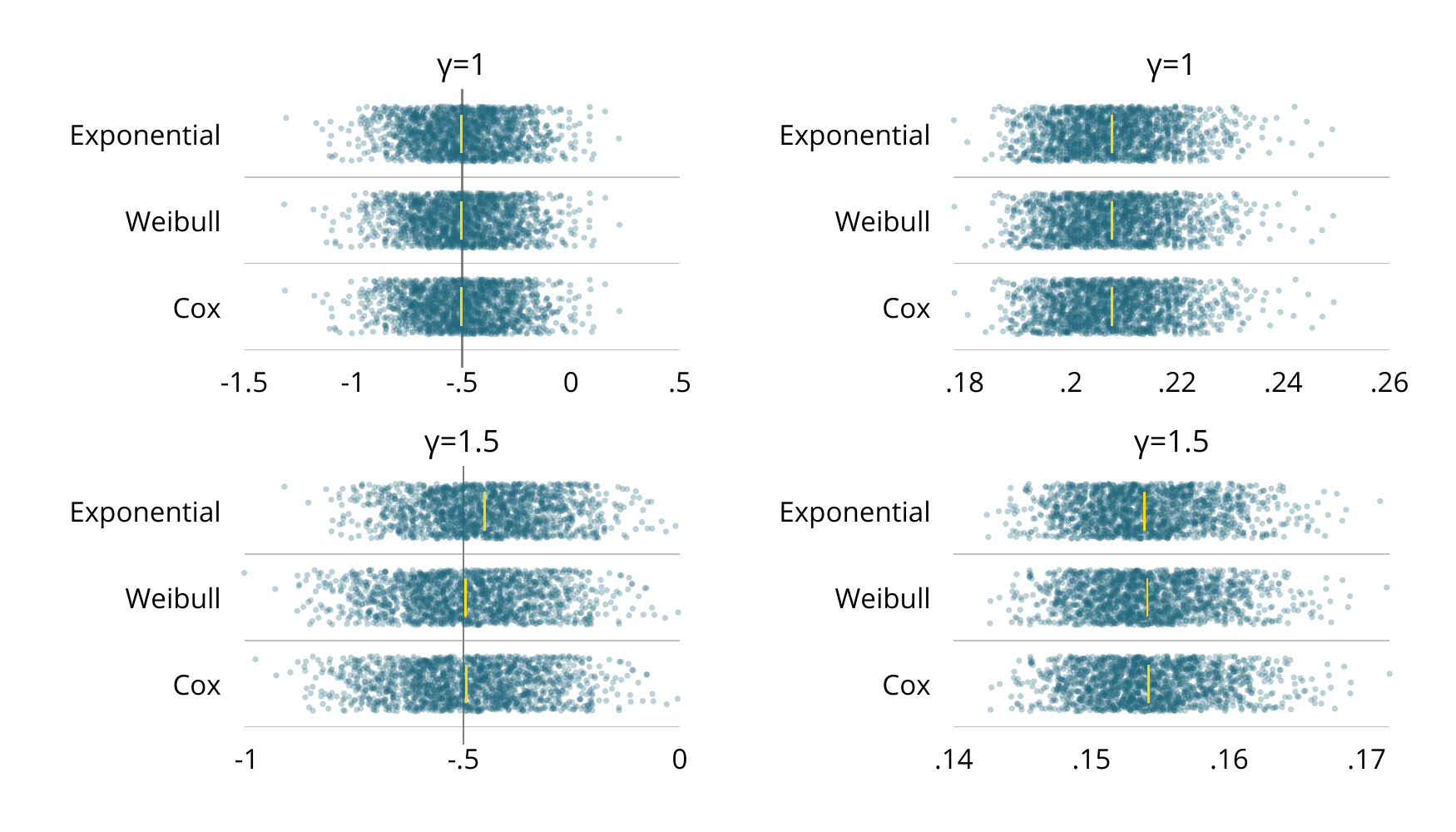} \\
\end{figure}

We first explore the raw results. Figure \ref{f:thetai} plots the estimates $\hat{\theta}_i$ and  $\widehat{\text{SE}}(\hat{\theta})_i$ for the two data-generating mechanisms and three methods, with means displayed as yellow pipes. The left panels plot $\hat{\theta}_i$. It is clear that, when $\gamma = 1$, the mean and variance of $\hat{\theta}_i$ is very similar for the three methods. The mean is close to the true value of $\theta = -0.5$ for all methods. When in truth $\gamma = 1.5$, the empirical SE is slightly higher for all methods (because there are fewer events among the 500 observations under this data-generating mechanism). The exponential proportional-hazards model is now misspecified and we observe a shift of the mean of $\hat{\theta}_i$ towards the null, indicating some bias. The right panels of figure \ref{f:thetai} plot the estimated standard errors $\widehat{\text{SE}}(\hat{\theta})_i$. These are smaller for the upper panel ($\gamma=1$) than the lower panel ($\gamma=1.5$) but there is little to choose between the methods.

\begin{figure} \caption{Comparison of estimates for methods when $\gamma=1.5$, where each point represents one repetition. Panel (A): Upper triangle displays $\hat{\theta}_i$; lower triangle displays $\widehat{\text{SE}}(\hat{\theta}_i)$. Panel (B): Plot of difference \textit{vs.}\ mean for $\hat{\theta}_i$ and $\widehat{\text{SE}}(\hat{\theta}_i)$, with Weibull as the comparator.\label{f:matrix}}
  \centering
  \includegraphics[width=.45\textwidth]{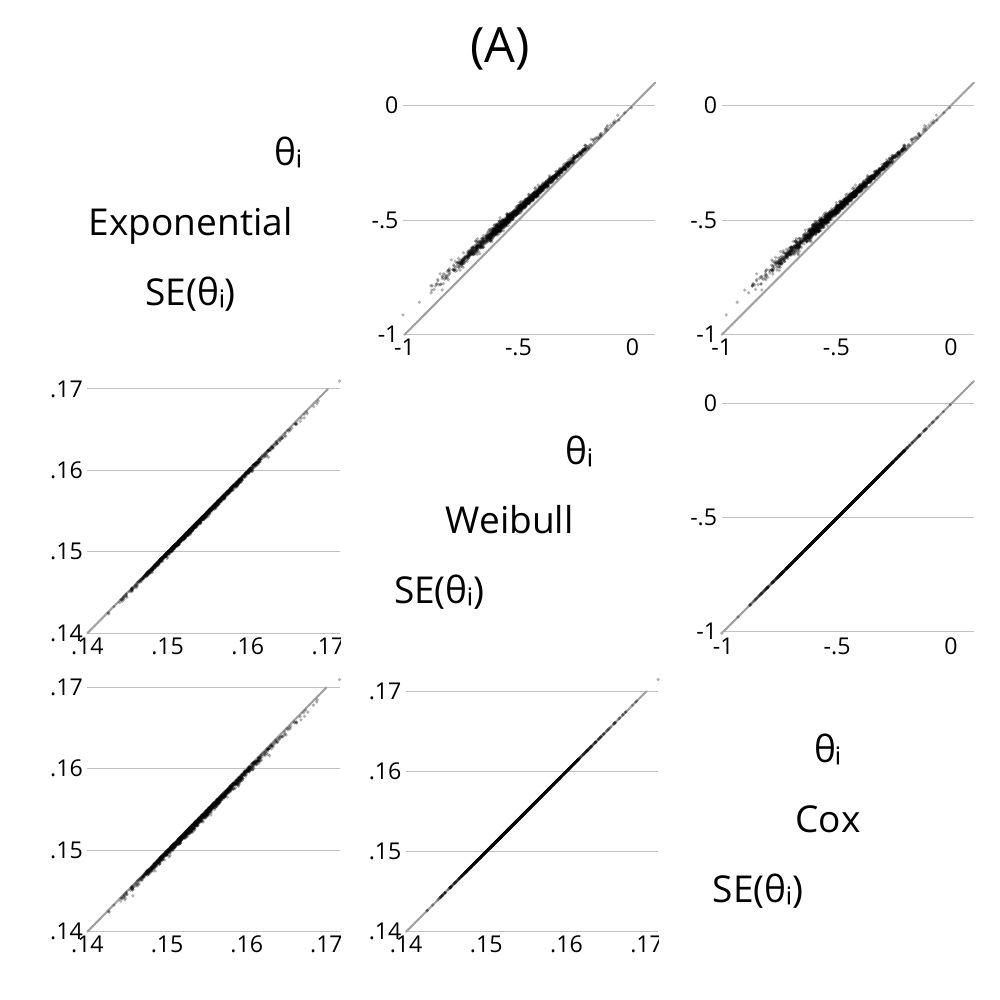}~
  \includegraphics[width=.45\textwidth]{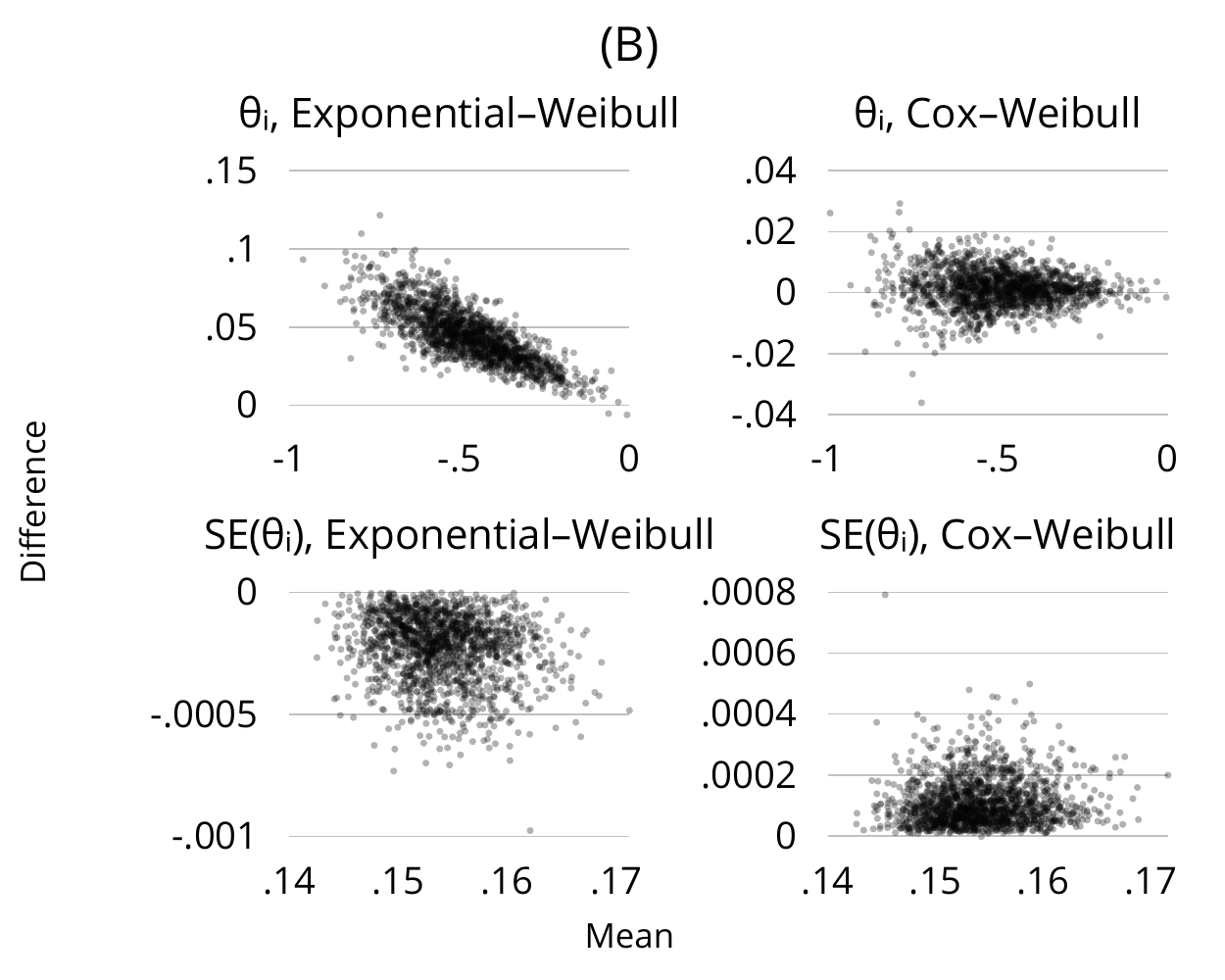}
\end{figure}

We next compare these estimates by plotting $\hat{\theta}_i$ for each method \textit{vs.}\ every other method, and the same for $\widehat{\text{SE}}(\hat{\theta})_i$. The data pairs come from the same repetition (\textit{i.e.}\ they are estimated in the same simulated dataset) and are compared to the line of equality. This is done in figure~\ref{f:matrix}(a), for the second data-generating mechanism only ($\gamma = 1.5$), of interest  because the exponential model is misspecified. We can see that the estimates of both $\hat{\theta}_i$ and $\widehat{\text{SE}}(\hat{\theta}_i)$ are highly correlated across all methods. The upper triangle of plots in figure \ref{f:matrix} shows that, while $\hat{\theta}_i$ is almost identical for the Weibull and Cox models, it tends to be systematically closer to 0 for the exponential model. The estimates of $\widehat{\text{SE}}(\hat{\theta}_i)$ show that again, the estimates are extremely similar for the Weibull and Cox models, and are very slightly larger for the exponential model. Figure~\ref{f:matrix}(B) gives the corresponding plots of the difference \textit{vs.}\ mean (\textit{Weibull} is the comparator method here as it is correctly specified).

\begin{figure} \caption{`Zip~plot' of the $1,600$ confidence intervals for each data-generating mechanism and analysis method. The vertical axis is the fractional centile of $|z|$ with $z=(\hat{\theta}_i-\theta)/\text{ModSE}$ associated with the confidence interval.\label{f:zipplot}}
  \centering
  \includegraphics[width=.7\textwidth]{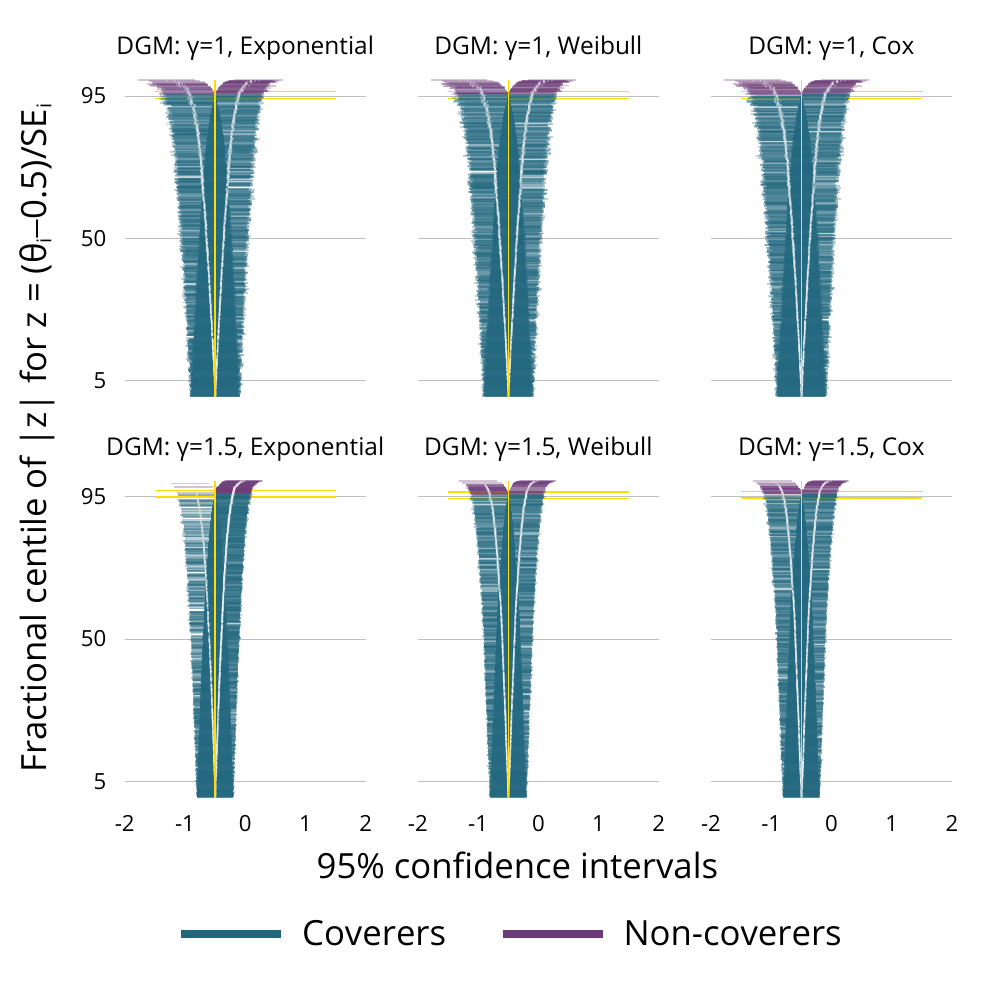} \\
\end{figure}

Figure \ref{f:zipplot} is a new visualisation, the `zip~plot', which helps to understand coverage by viewing the confidence intervals directly (implemented in Stata; for implementations in R and \textsc{Sas}, see \cite{rsimsum} and \cite{zipper18} respectively). For each data-generating mechanism and method, the confidence intervals are fractional-centile-ranked according to $|z|$, where $z=(\hat{\theta}_i-\theta)/\text{ModSE}$. This ranking is used for the vertical axis and is plotted against the intervals themselves. Intervals which cover $\theta$ are coloured blue (bottom); those which do not cover are coloured purple (top). When a method has 95\% coverage, the colour of the intervals switches at 95 on the vertical axis. The yellow horizontal lines are Monte Carlo 95\% confidence intervals for per~cent coverage. (As a general comment, note that the interesting area in many zip~plots will be near to the top and so it will often be more informative to `zoom~in' on the action, as suggested by Rick Wicklin in~\cite{zipper18}.)

In figure \ref{f:zipplot}, the upper panel again displays the results when $\gamma=1$ and the lower panel when $\gamma=1.5$. Despite coverage being approximately 95\% as advertised, there are more intervals to the right of $\theta=-0.5$ than to the left, particularly for those that do not cover $\theta$. This indicates that the model SEs must overestimate the empirical SE, because coverage is adequate despite bias. A zip~plot helps to make such a feature clear.

\subsection{Analysis of example} \label{ss:egresults}

The previous section demonstrated some exploratory analyses that may be of value. Next, we estimate performance for the measures of interest and present them in a table for which (we hope) the \textsc{ademp} structure is clear: different performance measures are stacked vertically; for each performance measure, the results for the two data-generating mechanisms each occupy one row; results for different methods are arranged across three columns (with Monte Carlo SEs in parentheses at a smaller point size than the estimate); there is only one estimand.

\begin{table} \centering
\caption{Estimates of performance for measures of interest {\footnotesize (Monte Carlo SEs in parentheses)}. Concerning results are highlighted in bold. See also figure~\ref{f:lolly}.\label{t:egpm}}
  \centering
  \begin{tabular}{ll|rlrlrl}
  \hline
  Performance & Data-generating & \multicolumn{6}{c}{Method} \\
  measure & mechanism & \multicolumn{2}{c}{Exponential} & \multicolumn{2}{c}{Weibull} & \multicolumn{2}{c}{Cox} \\
  \hline
  Bias & $\gamma=1$ & $-0.003$ & {\footnotesize (0.005)} & $-0.003$ & {\footnotesize (0.005)} & $-0.002$ & {\footnotesize (0.005)} \\
  & $\gamma=1.5$ & \textbf{0.049} & {\footnotesize (0.003)} & 0.005 & {\footnotesize (0.004)} & 0.006 & {\footnotesize (0.004)} \\
  \hline
  Coverage & $\gamma=1$ & 95.4\% & {\footnotesize (0.5)} & 95.4\% & {\footnotesize (0.5)} & 95.4\% & {\footnotesize (0.5)} \\
  & $\gamma=1.5$ & 96.0\% & {\footnotesize (0.5)} & 95.6\% & {\footnotesize (0.5)} & 95.8\% & {\footnotesize (0.5)}\\
  \hline
  Bias-eliminated & $\gamma=1$ & 95.6\% & {\footnotesize (0.5)} & 95.3\% & {\footnotesize (0.5)} & 95.4\% & {\footnotesize (0.5)} \\
  coverage & $\gamma=1.5$ & \textbf{97.2\% } & {\footnotesize (0.4)} & 95.7\% & {\footnotesize (0.5)} & 96.1\% & {\footnotesize (0.5)}\\
  \hline
  Empirical SE & $\gamma=1$ & 0.209 & {\footnotesize (0.004)} & 0.209 & {\footnotesize (0.004)} & 0.209 & {\footnotesize (0.004)} \\
  & $\gamma=1.5$ & 0.138 & {\footnotesize (0.002)} & 0.152 & {\footnotesize (0.003)} & 0.151 & {\footnotesize (0.003)} \\
  \hline
  Relative precision & $\gamma=1$ & 0.2\% & {\footnotesize (0.1)} & 0 & {\footnotesize (--)} & 0.3\% & {\footnotesize (0.1)} \\
  gain \textit{vs.}\ Weibull & $\gamma=1.5$ & 20.5\% & {\footnotesize (0.4)} & 0 & {\footnotesize (--)} & 0.6\% & {\footnotesize (0.2)} \\
  \hline
  Model SE & $\gamma=1$ & 0.208 & {\footnotesize (<0.001)} & 0.208 & {\footnotesize (<0.001)} & 0.208 & {\footnotesize (<0.001)} \\
  & $\gamma=1.5$ & 0.154 & {\footnotesize (<0.001)} & 0.154 & {\footnotesize (<0.001)} & 0.154 & {\footnotesize (<0.001)} \\
  \hline
  Relative error in & $\gamma=1$ & $-0.7$\% & {\footnotesize (1.8)} & $-0.7$\% & {\footnotesize (1.8)} & $-0.5$\% & {\footnotesize (1.8)} \\
  Model SE & $\gamma=1.5$ & \textbf{11.5\% } & {\footnotesize (2.0)} & 1.7\% & {\footnotesize (1.8)} & 2.1\% & {\footnotesize (1.8)} \\
  \hline
  \end{tabular}
\end{table}

\begin{figure} \caption{Lollipop plot of performance for measures of interest (Monte Carlo 95\% confidence intervals in parentheses). Concerning features need not be highlighted since they are readily visible. See also table \ref{t:egpm}.\label{f:lolly}}
  \centering
  \includegraphics[width=.7\textwidth]{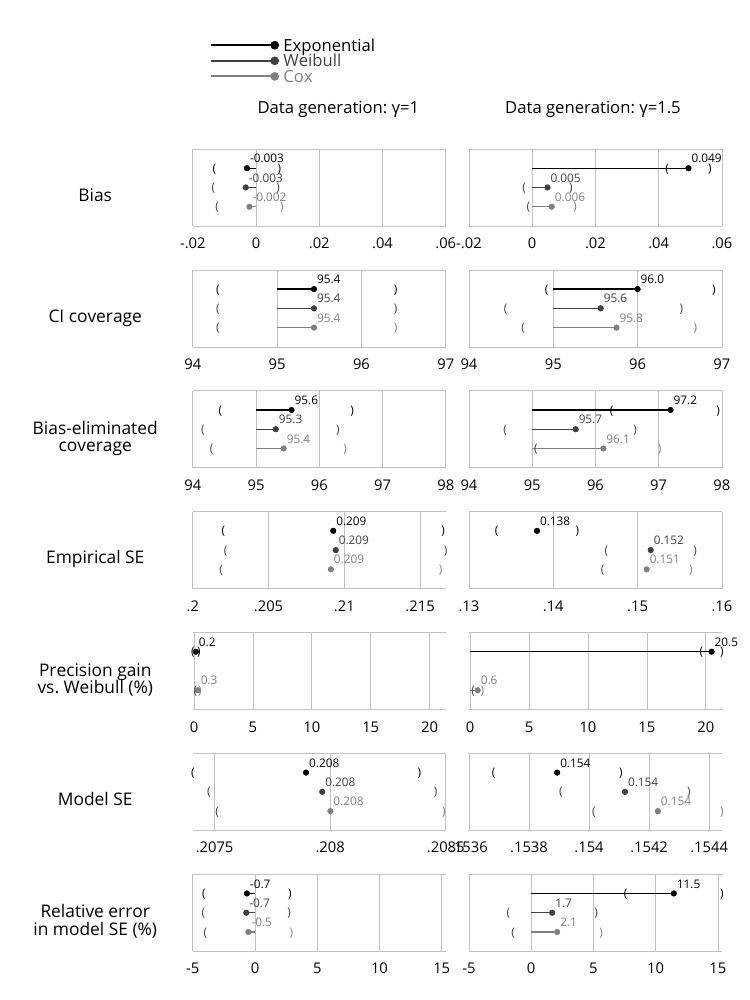} \\
\end{figure}

Also given in figure \ref{f:lolly} is an alternative graphical presentation of estimated performance called a \textit{lollipop plot}. The \textsc{ademp} structure is slightly different to the table but again clear: different performance measures are stacked vertically; for each performance measure the results for the three methods now occupy one row each; results for different methods are arranged across the two columns. Monte Carlo 95\% confidence intervals are now represented via parentheses (a visual cue due to the usual presentation of intervals as two numbers within parentheses).

The results confirm more formally some of the features we saw in our exploration of the estimates data. The interesting features concern the exponential model when $\gamma=1.5$, since the Weibull and Cox models behave well in all cases. We see that the exponential model suffers some bias towards the null, which is approximately 10\% of the true value. This is non-negligible. Next, we see that coverage is still over the nominal 95\%, which is surprising in the presence of bias. The empirical~SE is the same for all models when $\gamma=1$ and lowest for the exponential model when $\gamma=1.5$, while the Weibull and Cox models are very similar; recall however that in the presence of different biases, the empirical~SE is not comparable across methods. For relative precision (\textit{vs.} the Weibull model) a very similar pattern is seen as for empirical SE. The Model SE is the same for all methods and data-generating mechanisms. This explains why the exponential model has acceptable coverage when $\gamma=1.5$: the bias is cancelled out by the fact that the model SE is overestimated. This is confirmed by the relative error in Model SE.

Looking at table \ref{t:egpm}, the Monte Carlo SEs of performance estimates are all acceptable and so we would be happy to draw conclusions about the methods based on the 1,600 repetitions.

\subsection{Conclusions of example}
When an exponential model is misspecified, the hazard ratio can be biased. Probably not by much. Further research is needed.

More seriously, note that the data-generating mechanisms we used do not cover any real breadth of scenarios. For example, we might have explored varying $\no$, $\lambda$ and $\theta$ over a range of values to explore when issues are present.

\section{Concluding remarks} \label{s:conclusions}

Simulation studies are an invaluable tool for research into statistical methods, evidenced by the large proportion of Volume~34 \textit{Statistics in Medicine} articles whose conclusions relied in part on simulation studies. Because methods promoted may be used in medical research (or many other scientific areas), transparent reporting of the design and execution of simulation studies is critical.

While simulation studies are widely used, they tend to be poorly reported by those who publish their results.

There are many areas to be improved in the reporting of simulation studies. Our view is that the two main shortcomings are (i) lack of clarity over the design, which \textsc{ademp} aims to deal with, and (ii) failure to report estimates of Monte-Carlo uncertainty.

We have described -- and advocate -- a structured approach to the planning of simulation studies that involves identifying \textit{aims}, \textit{data-generating mechanisms}, \textit{methods}, \textit{estimands} and \textit{performance measures}. All of these and the rationale for decisions should be included in reporting. For an excellent example of a clearly described design, see Austin and Stuart\cite{austin15}. Reports of simulation studies are now beginning to explicitly use the \textsc{ademp} structure; see Thompson \textit{et al.}\cite{thompson17}, Sayers \textit{et al.}\cite{sayers17}, Morris \textit{et al.}\cite{morris17ma12}, Keogh \& Morris\cite{keogh18} and Pham \textit{et al.}\cite{pham18}.

We have given formulas for computing the Monte Carlo standard error for the most common performance measures, and made some suggestions about reporting. Note that the Stata package\texttt{ simsum}\cite{White10simsum} and R package{ rsimsum}\cite{rsimsum} automate this process for commonly used performance measures. See Boos and Osborne for more general assessment of Monte Carlo SEs for complex performance measures\cite{boos15}.

\subsection{Future directions} \label{s:future}

Three areas that we regard as of increasing future importance are simulation protocols, release of code, and consortia of authors. We discuss these as a step towards resolving occurrences of simulation studies with contradictory results.

A charitable view of such contradictory results, which we tend to hold, is that methods are developed by researchers who concerned with handling the specific problems they have seen in practice. Given such a background, they are running the relevant simulation studies (that may not be relevant to others). A less charitable possibility might be selective reporting of only the most favourable (or unfavourable) configurations of data-generating mechanisms, running the simulations many times under different seeds and selecting the most favourable\cite{grieve16}, and more (left to the reader's imagination).

A starting point to addressing this issue is to write detailed simulation protocols before writing code. This would ideally protect against authors choosing to report favourable results and force one to be clear about `\textsc{ademp}' in advance. Of course, the weak point is that simulation studies do not to need approval before `data collection', so protocols could be written after-the-fact and this cannot be clear even with published protocols. The counterargument is that protocols must justify the rationale for choices: as well as describing \textit{what} is planned, there is a burden to explain \textit{why}.

Due to the prejudices introduced by experiences of data, it is constructive for authors who have produced contradictory simulation results to work together on `late-phase' simulation studies (using an analogy from clinical trials in drug development). This allows robust discussion of the design and exploration of disparities among previous work. One exemplar of such an approach is in methods for handling incomplete data where the analysis has a multilevel structure. Three groups of researchers had developed methods and worked together on understanding how the methods differ and on simulation studies to evaluate their performance, resulting in the paper by Audigier \textit{et~al}\cite{audigier18}. This approach is in our view more satisfactory than a group of researchers executing a large, late-phase simulation study without the input of authors of previous work, though this strategy is sometimes adopted\cite{langan17}.

Boulestix, Wilson and Hapfelmeier raise an important consideration \cite{boulestix17} for benchmarking studies that is also relevant to late-phase simulation studies: it is easy to assume that the performance of methods is due entirely to the methods themselves, but this ignores the fact that the implementation of a method involves the knowledge and skill of the individuals involved. A poor implementation may make a superior method appear weak. This suggests again that including consortia of authors, as in Audigier \textit{et~al.}, is advisable.

No simulation study is definitive and new methods or refinements of methods are inevitable. For researchers wishing to replicate or extend the results of earlier simulation studies, the design of earlier work must have been written out fully and unambiguously. This can be a difficult task and, to ensure this, authors should release simulation code publicly (a policy that is now encouraged by some journals, notably \textit{Biometrical Journal}, and required by others). The happy corollary is that code may be checked more thoroughly by its authors if it is subject to external scrutiny. There is generally no excuse for withholding code. One caveat is for resampling studies, where permissions to release the original data may be lacking (note that code for running the simulation can still be made available even if it cannot be run on the same data).

\subsection{Final remark}
Simulation studies are a powerful tool. However, it is important to be aware that, because a simulation study took hard work and thought, we are liable to believe it tells us more than it truly does. To quote Patrick Royston, `Simulation studies reveal points of light on a landscape, but can not illuminate the entire landscape.' We can hope and plan to illuminate important points and to build up a picture of the landscape, particularly where terrain may be particularly rocky or particularly fertile.

We hope that the guidance in this tutorial will improve researchers' understanding, planning, execution and future reporting of simulation studies.

\section*{Conflicts of interest}
All authors declare that they developed, and regularly deliver, a short course on simulation studies, from which this work grew, and from which they have benefited financially.

\section*{Acknowledgements}
Tim Morris and Ian White are supported by the Medical Research Council (grant numbers MC\_UU\_12023/21 and MC\_UU\_12023/29). Michael Crowther is partly supported by a Medical Research Council New Investigator Research Grant (grant number MR/P015433/1).

For thought-provoking discussions and input to this work, we thank Alessandro Gasparini, Tra Pham, Brennan Kahan, Ruth Keogh, Cl\'{e}mence Leyrat, Kristian Brock, Christian Hennig and Patrick Royston. We also thank the many participants who have attended our courses and whose questions and feedback provided the motivation for this article.

The initial manuscript and revised version of this article were released as pre-prints, and we are grateful to the people who contributed reviews and comments on these. In particular, those who made important substantive comments: Maarten van Smeden, Rolf Groenwold, Bas Penning-de Vries, Kim Luijken, Martina McMenamin, Paula Dhiman, Leane McCabe, Alessandro Gasparini, Stephen Senn, Adrian Sayers, Richard Torkar, David Mannheim, Daniel Oberski, Teague Henry and Rick Wicklin.

\bibliography{C:/bib/tpmorris}

\appendix

\section{Review: further information}\label{app:dgms}
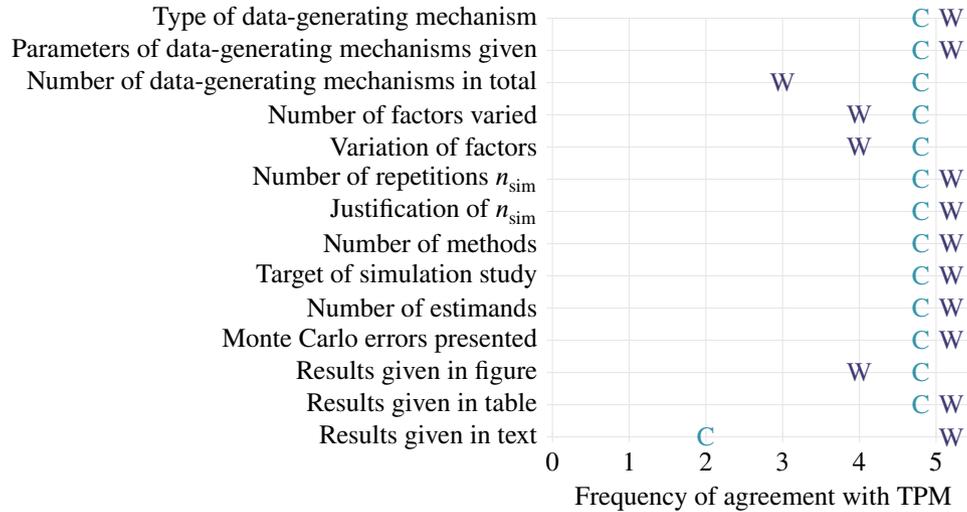
\begin{figure}
  \caption{Reviewer agreement on key variables for \textit{Statistics in Medicine} Volume~34 review. Frequency of agreement of TPM with IRW (marker \textcolor{v3}{W}) and MJC (marker \textcolor{v6}{C}). For the same frequency, \textcolor{v6}{C} is nudged left and \textcolor{v3}{W} right to avoid visual clash.
  \label{f:agree}}
  \centering
  \begin{tikzpicture}
    \centering
    \begin{axis}
      [
        width=.4\textwidth,
        axis x line=bottom,
        axis y line=left,
        axis line style={draw=none},
        xmin=0, xmax=5.5,
        ymajorgrids,
        xmajorgrids,
        grid style={draw=black!10},
        y=1.2em,
        tick style={draw=black!10},
        ytick=data,
        xtick={0,1,2,3,4,5},
        scaled ticks=false,
        yticklabels from table={\Ian}{0},
        xticklabel style={/pgf/number format/fixed},
        xlabel=Frequency of agreement with TPM
      ]
      \addplot[v3,only marks,mark=text,text mark=W] table [
          y expr=-\coordindex,
          x index=1
      ]{\Ian};
      \addplot[opacity=.8,v6,only marks,mark=text,text mark=C] table [
          y expr=-\coordindex,
          x index=1
      ]{\Michael};
    \end{axis}
  \end{tikzpicture}
\end{figure}

Agreement between TPM and each of the other authors was measured on 14 key variables for the review (IRW and MJC did not review any of the same articles). The variable-by-variable results are given in figure \ref{f:agree}. Note that for some variables, there were two possible choices (e.g.\ `Results given in a figure Y/N') while for others there were many (e.g.\ `Number of repetitions $\ns$'). Overall, TPM agreed with the other reviewers in 132 of 140 answers (94\%). IRW and TPM agreed on 65 of 70 answers (93\%); MJC and TPM agreed on 67 or 70 answers (96\%).

\pgfplotstableread[header=false,col sep=comma]{
  Type of data-generating mechanism,5.2
  Parameters of data-generating mechanisms given,5.2
  Number of data-generating mechanisms in total,3
  Number of factors varied,4
  Variation of factors,4
  Number of repetitions $\ns$,5.2
  Justification of $\ns$,5.2
  Number of methods,5.2
  Target of simulation study,5.2
  Number of estimands,5.2
  Monte Carlo errors presented,5.2
  Results given in figure,4
  Results given in table,5.2
  Results given in text,5.2
} \Ian
\pgfplotstableread[header=false,col sep=comma]{
  Type of data-generating mechanism,4.8
  Parameters of data-generating mechanism/s given,4.8
  Number of data-generating mechanisms in total,4.8
  Number of factors varied,4.8
  Variation of factors,4.8
  Number of repetitions $n_{sim}$,4.8
  Justification of $n_{sim}$,4.8
  Number of methods,4.8
  Target of simulation study,4.8
  Number of estimands,4.8
  Monte Carlo errors presented,4.8
  Results given in figure,4.8
  Results given in table,4.8
  Results given in text,2
} \Michael

Figure \ref{f:dgms} gives summary information about how data were generated. Panel (a) shows that there was great variation in the total number of data-generating mechanisms, with the majority of simulation studies using under 20, and the largest number being $\sim 41$~billion. Panel (b) shows that simulation studies tended to vary few factors (with one exception). For the simulation studies varying more than one factor, the most common way to do this was in a fully factorial manner (panel (c)). However, some studies varied the factors one-at-a-time and others mixed the two together. Unfortunately, not all simulation studies noted the number of repetitions (panel (d)). The most common choices of $\ns$ were, in descending order: $1,000$, $500$ and $10,000$.

\pgfplotstableread[header=false,col sep=semicolon]{
  1;5
  2;3
  3;12
  4;10
  5;1
  6;4
  7;2
  9;2
  10;3
  12;8
  14;1
  16;3
  18;6
  20;2
  24;3
  26;1
  27;2
  28;1
  29;1
  30;1
  32;1
  36;3
  52;1
  54;3
  55;1
  56;1
  60;1
  72;1
  81;1
  84;1
  90;1
  96;2
  108;4
  144;1
  162;1
  168;1
  192;1
  288;1
  9,090;1
  9,600;1
  4.184e+10;1
}\ndgms
\pgfplotstableread[header=false,col sep=comma]{
  0,5
  1,13
  2,29
  3,22
  4,15
  5,5
  6,5
  7,2
  8,1
  9,1
  14,1
  324,1
}\fvaried
\pgfplotstableread[header=false,col sep=comma]{
  Factorial,57
  N/A,19
  One-at-a-time,8
  Other,4
  Partly factorial,12
}\howvaried
\pgfplotstableread[header=false,col sep=semicolon]{
Not stated;4
1;1
50;3
100;3
200;3
400;1
500;14
1,000;41
2,000;5
3,000;3
5,000;5
10,000;10
20,000;1
100,000;2
1,000,000;4
}\nsim
\begin{figure}
  \caption{Results of \textit{Statistics in Medicine} Volume~34 review for data-generating mechanisms. Values are both frequency and \% 
  \label{f:dgms}}
  \begin{tabular}[t]{rr}
  \mbox{
  \begin{tikzpicture}
    \centering
    \begin{axis}
      [
        width=.35\textwidth,
        axis x line=bottom,
        axis y line=left,
        axis line style={draw=none},
        xmin=0, xmax=13,
        ymajorgrids,
        xmajorgrids,
        grid style={draw=black!10},
        y=1.2em,
        tick style={draw=black!10},
        ytick=data,
        scaled ticks=false,
        yticklabels from table={\ndgms}{0},
        xticklabel style={/pgf/number format/fixed},
        xlabel=Frequency (and \%),
        title=a) Total number of data-generating mechanisms
      ]
      \addplot[only marks, mark options={fill=white}] table [
          y expr=-\coordindex,
          x index=1
      ]{\ndgms};
    \end{axis}
  \end{tikzpicture}
  }
  &
  \mbox{
  \begin{tabular}[b]{r}
  \begin{tikzpicture}
    \centering
    \begin{axis}
      [
        width=.35\textwidth,
        axis x line=bottom,
        axis y line=left,
        axis line style={draw=none},
        xmin=0, xmax=30,
        ymajorgrids,
        xmajorgrids,
        grid style={draw=black!10},
        y=1.2em,
        tick style={draw=black!10},
        ytick=data,
        scaled ticks=false,
        yticklabels from table={\fvaried}{0},
        xticklabel style={/pgf/number format/fixed},
        xlabel=Frequency (and \%),
        title=b) Number of factors varied
      ]
      \addplot[only marks, mark options={fill=white}] table [
          y expr=-\coordindex,
          x index=1
      ]{\fvaried};
    \end{axis}
  \end{tikzpicture}
  \\
  \begin{tikzpicture}
    \centering
    \begin{axis}
      [
        width=.35\textwidth,
        axis x line=bottom,
        axis y line=left,
        axis line style={draw=none},
        xmin=0, xmax=60,
        ymajorgrids,
        xmajorgrids,
        grid style={draw=black!10},
        y=1.2em,
        tick style={draw=black!10},
        ytick=data,
        scaled ticks=false,
        yticklabels from table={\howvaried}{0},
        xticklabel style={/pgf/number format/fixed},
        xlabel=Frequency (and \%),
        title=c) How factors were varied
      ]
      \addplot[only marks, mark options={fill=white}] table [
          y expr=-\coordindex,
          x index=1
      ]{\howvaried};
    \end{axis}
  \end{tikzpicture}
  \\
  \begin{tikzpicture}
    \centering
    \begin{axis}
      [
        width=.35\textwidth,
        axis x line=bottom,
        axis y line=left,
        axis line style={draw=none},
        xmin=0, xmax=45,
        ymajorgrids,
        xmajorgrids,
        grid style={draw=black!10},
        y=1.2em,
        tick style={draw=black!10},
        ytick=data,
        scaled ticks=false,
        yticklabels from table={\nsim}{0},
        xticklabel style={/pgf/number format/fixed},
        xlabel=Frequency (and \%) ,
        title=d) $\ns$ per data-generating mechanism
      ]
      \addplot[only marks, mark options={fill=white}] table [
          y expr=-\coordindex,
          x index=1
      ]{\nsim};
    \end{axis}
  \end{tikzpicture}
  \end{tabular}
  }
  \end{tabular}
\end{figure}

\begin{figure}
  \caption{Results of \textit{Statistics in Medicine} Volume~34 review for estimands (a) and methods (b) evaluated.\label{f:em}}
  \centering
  \begin{tabular}[t]{rr}
  \begin{tikzpicture}
    \centering
    \begin{axis}
      [
        width=.35\textwidth,
        axis x line=bottom,
        axis y line=left,
        axis line style={draw=none},
        xmin=0, xmax=35,
        ymajorgrids,
        xmajorgrids,
        grid style={draw=black!10},
        y=1.2em,
        tick style={draw=black!10},
        ytick=data,
        scaled ticks=false,
        yticklabels from table={\nestimands}{0},
        xticklabel style={/pgf/number format/fixed},
        xlabel=Frequency (and \%) ,
        title=a) Number of estimands
      ]
      \addplot[only marks, mark options={fill=white}] table [
          y expr=-\coordindex,
          x index=1
      ]{\nestimands};
    \end{axis}
  \end{tikzpicture}
  &
  \begin{tikzpicture}
    \centering
    \begin{axis}
      [
        width=.35\textwidth,
        axis x line=bottom,
        axis y line=left,
        axis line style={draw=none},
        xmin=0, xmax=25,
        ymajorgrids,
        xmajorgrids,
        grid style={draw=black!10},
        y=1.2em,
        tick style={draw=black!10},
        ytick=data,
        scaled ticks=false,
        yticklabels from table={\nmethods}{0},
        xticklabel style={/pgf/number format/fixed},
        xlabel=Frequency (and \%) ,
        title=b) Number of methods evaluated
      ]
      \addplot[only marks, mark options={fill=white}] table [
          y expr=-\coordindex,
          x index=1
      ]{\nmethods};
    \end{axis}
  \end{tikzpicture}
  \end{tabular}
\end{figure}
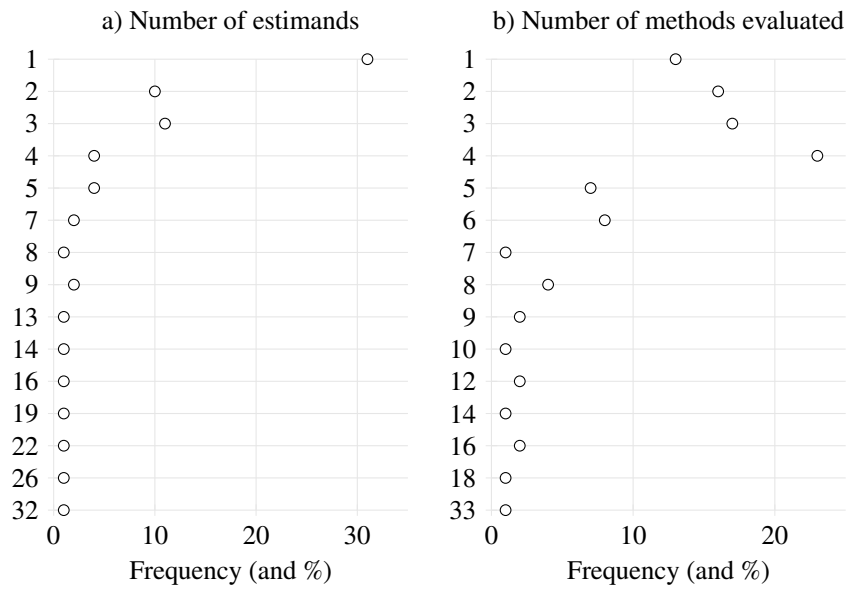

Figure~\ref{f:em}a shows the number of estimands evaluated by the simulation studies included in the review. In general, there were few, with a single estimand the most common. Figure~\ref{f:em} panel (b) gives the number of methods evaluated by the simulation studies included in the review (right panel). The majority evaluated few methods (with four the most common number). This suggests that simulation studies provide a proof-of-concept, or that the methods are designed for new problems for which there are few alternatives available.

\pgfplotstableread[header=false,col sep=comma]{
1,31
2,10
3,11
4,4
5,4
7,2
8,1
9,2
13,1
14,1
16,1
19,1
22,1
26,1
32,1
}\nestimands
\pgfplotstableread[header=false,col sep=comma]{
1,13
2,16
3,17
4,23
5,7
6,8
7,1
8,4
9,2
10,1
12,2
14,1
16,2
18,1
33,1
}\nmethods

Table~\ref{t:software} lists the software packages mentioned and the number of mentions in simulation studies included in the review. This was based on a lenient judgement: for example, many articles mentioned a software package in which a method was implemented but did not mention what software was used to run the simulation study.

\begin{table} \centering
\caption{Software mentioned in simulation reports, review of \textit{Statistics in Medicine} Volume~34. Note that there are more than 100 entries as some articles reported more than one package.\label{t:software}}
\begin{tabular}{ll}
  \hline
  Software & Freq.\  \\
  \hline
  \textit{None mentioned} & 38 \\
  C & 1 \\
  \textsc{Jags} & 1 \\
  \textsc{Matlab} & 1 \\
  R & 41 \\
  \textsc{Sas} & 17 \\
  Sa\textsc{ts}can & 1 \\
  Stata & 4 \\
  Stat\textsc{x}act & 1 \\
  Win\textsc{bugs} & 3 \\
  \hline
\end{tabular} \\
\end{table}

\end{document}